\documentclass[10pt]{iopart}
\expandafter\let\csname equation*\endcsname\relax
\expandafter\let\csname endequation*\endcsname\relax
\usepackage{graphicx}
\usepackage{dcolumn}
\usepackage{makecell}
\usepackage{mathtools}
\usepackage{array}
\usepackage{multirow}
\usepackage{diagbox}
\usepackage{fp}
\usepackage{pict2e}
\usepackage{keyval}
\usepackage{calc}
\usepackage{comment}
\usepackage{soul}
\usepackage{iopams} 
\usepackage[colorlinks=true, allcolors=blue]{hyperref}
\usepackage{cite}
\usepackage{svg} 
\usepackage{orcidlink}

\usepackage[algo2e]{algorithm2e} 
\usepackage{algorithm}

\expandafter\let\csname equation*\endcsname\relax
\expandafter\let\csname endequation*\endcsname\relax
\usepackage{amsmath}

\newcommand{\R}{\mathbb{R}}
\newcommand{\s}{\mathcal{S}}

\begin{document}

\title[Inferring traits of hyperuniformity from local structures via persistent homology]
{Inferring traits of hyperuniformity from local structures via persistent homology}

\author{Abel H. G. Milor$^1$ \orcidlink{0009-0001-3890-0608} and Marco Salvalaglio$^{1,2}$ \orcidlink{0000-0002-4217-0951}}
\address{$^1$Institute of Scientific Computing, TU Dresden, 01062 Dresden, Germany.}
\address{$^2$Dresden Center for Computational Materials Science, TU Dresden, 01062 Dresden, Germany.}
\ead{abel\_henri\_guillaume.milor@tu-dresden.de}
\ead{marco.salvalaglio@tu-dresden.de}

\begin{abstract}
Hyperuniformity refers to the suppression of density fluctuations at large scales. Typical for ordered systems, this property also emerges in several disordered physical and biological systems, where it is particularly relevant to understand mechanisms of pattern formation and to exploit peculiar attributes, e.g., interaction with light and transport phenomena. While hyperuniformity is a global property, ideally defined for infinitely extended systems, several disordered correlated systems have finite size. It has been shown in [\href{http://www.doi.org/10.1103/PhysRevResearch.6.023107}{Phys. Rev. Research 6, 023107 (2024)}] that global hyperuniform characteristics systematically correlate with distributions of topological properties representative of local arrangements. In this work, building on this information, we explore and assess the inverse relationship between hyperuniformity and local structures in point patterns as described by persistent homology. Standard machine learning algorithms trained on persistence diagrams are shown to detect hyperuniformity of periodic point patterns with high accuracy. Therefore, we demonstrate that the information on patterns' local structures allows for characterizing whether finite size arrangements are analogous to those realized in hyperuniform patterns.
Then, addressing more quantitative aspects, we show that parameters defining hyperuniformity globally can be reconstructed by comparing persistence diagrams of targeted patterns with reference ones. We also explore the generation of patterns entailing given topological properties. The results of this study pave the way for advanced analysis of hyperuniform patterns including local information, and introduce basic concepts for their inverse design.
\end{abstract}

\vspace{2pc}

\ioptwocol

\section{Introduction}

Density fluctuations are fundamental to many-particle systems, influencing properties from material strength to biological organization. In certain systems, these fluctuations exhibit a peculiar behavior known as \textit{hyperuniformity}: they significantly diminish or vanish entirely at large scales \cite{torquato2003local,torquato2018hyperuniform}. This property is present in ordered systems like crystals. However, disordered systems can also be hyperuniform, and they are of particular interest because they share some of the effective properties at large scales with ordered systems. Disordered hyperuniform (HU) systems emerge in different contexts, e.g., from cosmology \cite{Gabrielli2002,GabrielliPRD2003} to biology \cite{JiaoPRE2014,mayer2015well,Backofen24} and condensed matter \cite{xie2013hyperuniformity,martelli2017largescale,salvalaglio2020hyperuniform}, making their study broadly relevant for understanding structural organization and pattern formation.

Controlling the design of HU arrangements has also proved useful to tailoring properties in resilient environments as shown, for instance, for some applications in optics \cite{torquato2018hyperuniform,yu2021engineered,Vynck2023}. In this context, generating hyperuniform patterns is crucial, and several methods have been proposed \cite{Uche2004,Uche2006,florescu2009designer,Morse2023,Aaron2024}. The inverse design of HU structures given a targeted property has been explored less. Recently, an approach to inverse design HU, typical non-HU and anti-HU patterns by tuning their pair statistics has been proposed \cite{Wang2024}. An interesting aspect of this task consists of the existence of manifolds of HU patterns that are similar in terms of long-range decrease of density fluctuations while exhibiting different arrangements on small scales or, vice versa, they are indistinguishable by the naked eye but globally exhibit different behaviors \cite{torquato2018hyperuniform}. 

Given its generality, hyperuniformity has been widely investigated for arrangements of points and fields of different kinds \cite{torquato2018hyperuniform,ma2017random}. It is usually defined in terms of global quantities, such as the number variance sampled over large regions or, equivalently, the limit of the structure factor for vanishing wave numbers. However, although hyperuniformity is usually quantified in terms of global characteristics, it also affects arrangements on relatively small scales. In addition, physical systems that may exhibit hyperuniformity are very often realized over finite-size domains. This is, for instance, the case of metamaterials for mechanical applications leveraging spinodal patterns \cite{Kumar2020}. It may also result from the self-assembly of HU patterns \cite{salvalaglio2020hyperuniform} when confined to prescribed finite-size regions (e.g., via patterning), and could emerge in HU active systems \cite{Zheng2024} which are often confined to one phase in two- or multi-phase systems \cite{Klamser2018}. Descriptions beyond classical analysis of long-wavelength behavior are thus in general needed to fully characterize these systems.

Correlations between hyperuniformity and distributions of local structures have been recently shown in reference~\cite{salvalaglio2024persistent} with the analysis of the statistics of local (graph) neighborhood motifs \cite{Skinner2021,skinner2022topological} of HU and non-HU point patterns, as well as using an up-and-coming technique which allows the characterization of the ``shape" of data sets, the so-called \textit{persistent homology} \cite{edelsbrunner2002topological, edelsbrunner2008persistent, edelsbrunner2022computational}. The latter, exploited in this work too, is a topological data analysis method that tracks the appearance (\textit{birth}) and disappearance (\textit{death}) of homological features (such as connected components, loops, and voids) upon a given parametrization of patterns (e.g., a threshold for scalar fields) and is further detailed below. Importantly, persistent homology analyzes patterns via a parametrization that echoes characterizations and constructions already discussed in the context of hyperuniformity for two-phase media, such as the decoration of points with circles \cite{Torquato2002random,torquato2018hyperuniform} and the concept of spreadability of a solute in a surrounding matrix~\cite{torquato2021diffusion}.

Statistics of local neighborhoods and persistent homology applied to HU patterns unveiled crucial evidence. 
For instance, different subsets of an ideal HU pattern --resembling realistic conditions realized in a finite-size and non-periodic sampling of experimental systems or numerical simulations-- have different structure factors but similar distributions of topological properties. These may be then used for their detection and inverse design. However, except first attempts \cite{Wang2024}, the latter is yet to be addressed in detail. 

This work keeps focusing on the correspondence between hyperuniformity and topological features of HU point patterns first disclosed in reference~\cite{salvalaglio2024persistent}. In particular, given the \textit{direct connection} between global HU character and local arrangements discussed therein, it analyzes the corresponding \textit{inverse relations}. In other words, we inspect if and how hyperuniformity and specific HU characters can be devised by looking at local arrangements characterized by persistent homology. More in general, we aim to determine if geometrical arrangements in a given pattern are similar to those realized in ideal HU patterns.
Persistent homology involves data sets conveniently represented in the so-called \textit{persistence diagrams}, which cannot be inverted straightforwardly. However, remarkable methods have been proposed that consider representative \textit{features}, combining the information retained in these persistence diagrams and their exploitation in machine learning (ML) models \cite{pun2022machine_learning_persistence}. As shown below, this enables devising methods to determine whether a given persistent homology corresponds to a HU point pattern with high accuracy. Moreover, quantifying the (Wasserstein) distance between persistence diagrams with respect to reference ones allows us to estimate the effective structure factor parametrization. The inverse problem of reconstructing patterns from persistent homology is also explored, and we propose a method to achieve it in an approximate manner.

The manuscript is organized as follows. In section~\ref{sec:hu}, we recall the notion of hyperuniformity together with the essential quantities used in this work, as well as details concerning the generation of ideal HU patterns. Section~\ref{sec:persitent-homology} reviews the concept of persistent homology and the related persistence diagrams used to characterize the local arrangements in HU (and non-HU) patterns. The detection of HU or non-HU characters via ML models is reported in section~\ref{sec:machinelearning}. In section~\ref{sec:compute_parameters}, we illustrate the estimation of effective structure-factor parametrization by leveraging comparisons between persistence diagrams with reference ones. Section~\ref{sec:boundaries} additionally shows that the information contained in persistence diagrams can be exploited to detect boundaries between different realizations of nominally HU patterns. We finally report on the construction of point patterns to encode specific information about their geometry (persistence diagrams) in section~\ref{sec:reconstruction}. The main conclusions are discussed and summarized in section~\ref{sec:conclusions}. The data
and scripts used in this study are available at~\cite{repo}.

\begin{figure*}[h]
    \centering
    \includegraphics[width=\linewidth]{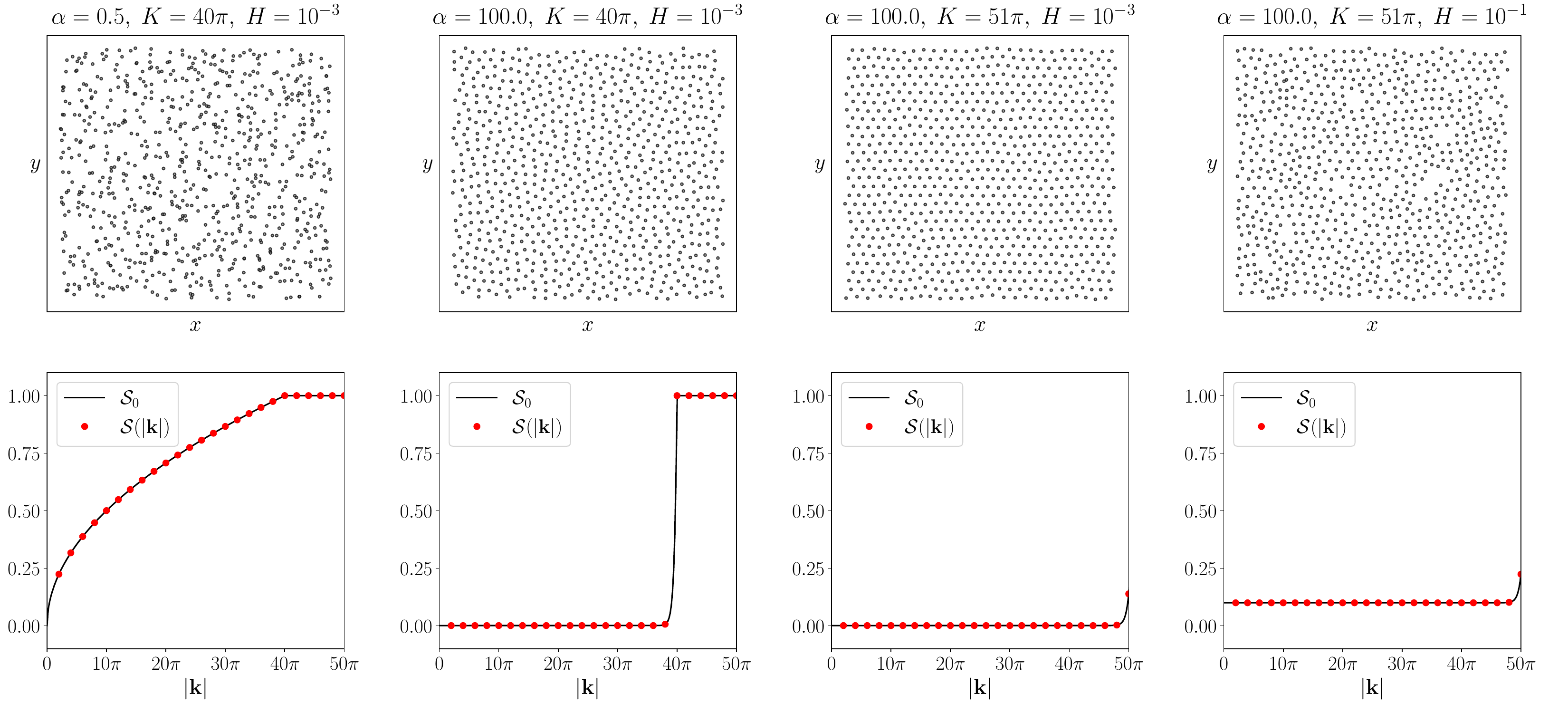}
    \vspace{-15pt}
    \caption{Examples of point patterns with a structure factor as in equation~\eqref{eq:formulaofstruc}. First row: Point patterns for different structure factor parameters reported above the panels. Second row: Structure factors, illustrated by both the $\s_0(|\mathbf{k}|)$ imposed in the generation procedure (solid black lines) and discrete values computed for the patterns in the first rows for $\s(|\mathbf{k}|)$ imposed at appropriate wavenumbers $|\mathbf{k}|=2\pi n$ with $n\in \mathbb{N}_{>0}$ (red dots). We remark that $\widetilde{H}$ in equation~\eqref{eq:htilde} corresponds to the value of $\s(|\mathbf{k}|)$ for the $|\mathbf{k}|=2\pi$, i.e. the leftmost red point in the structure factor plots and may deviate significantly from the parameter $H$ for relatively small $\alpha$. See, for instance, the case with $\alpha=0.5$ (first column).}
    \label{fig:points_Sk}
\end{figure*}

\section{Hyperuniformity}\label{sec:hu}

Consider a set $\mathcal{P}$ of points within the Euclidean space $\R^n$. For a given spherical observation window with radius $R$, namely a ball $\mathcal{B}_R$, we can define the number variance $\sigma^2(R)\,$=$\,\langle \mathcal{N}(R)^2 \rangle-\langle \mathcal{N}(R) \rangle^2$ with $\mathcal{N}(R)$ the number of points within $\mathcal{B}_R$. $\mathcal{P}$ is said to be HU if $\sigma^2(R)$ increases slower than the volume of $\mathcal{B}_R$ \cite{torquato2018hyperuniform}, i.e.:
\begin{equation}\label{eq:HUdef1}
    \lim\limits_{R\rightarrow \infty} \frac{\sigma^2(R)}{R^n} = 0.
\end{equation}
Intuitively, this can be understood as vanishing density fluctuations at large scales.
Equivalently, hyperuniformity can be defined via the limit for small wave numbers of the structure factor $\s(\mathbf{k})$,
\begin{equation}\label{eq:HUdef2}
    \lim\limits_{|\mathbf{k}|\rightarrow 0} \s(\mathbf{k}) = 0,
\end{equation}
where $\s(\mathbf{k})$ can be conveniently expressed as \cite{torquato2018hyperuniform}
\begin{equation}\label{eq:structurefactor}
    \s(\mathbf{k})= \frac{|\sum^{N}_{j=1} \text{exp}(i\mathbf{k}\cdot \mathbf{p}_j)|^2}{N}, \qquad |\mathbf{k}|\neq 0
\end{equation}
with $\mathbf{p}_j$ the coordinates of the $j$-th point in $\mathcal{P}$, $i$ the imaginary unit, and $N$ the number of points within $\mathcal{P}$. We remark that the definitions reported in equations~\eqref{eq:HUdef1}-\eqref{eq:HUdef2} formally hold for sets of infinite points. Focusing on the description via $\s(\mathbf{k})$, finite sets of points can be characterized by evaluating equation~\eqref{eq:structurefactor} up to the smallest $\mathbf{k}$ in the considered (finite) domain. Also, for rectangular domains spanned between the vectors $\{\mathbf{v}_i\}$, appropriate wave vectors for the evaluation of $\s(\mathbf{k})$ satisfy the condition $\mathbf{k}\cdot\mathbf{v}_i\,$=$\,2\pi n$, $n\in \mathbb{Z}$. For unitary vectors $\{\mathbf{v}_i\}$, this reduces to $\mathbf{k}\,$=$\,(2\pi n, 2\pi m)\neq \mathbf{0} $ with $n, m\in \mathbb{N}$.

In this work, we consider ideal patterns with a prescribed HU character. A point pattern with a target structure factor $\s_0$ can be obtained by minimizing the following objective function
\cite{Uche2004, Uche2006}
\begin{equation}\label{eq:tominimize}
    O(\{\mathbf p_j\})=\sum_{\mathbf{k}\in Q}|\s(\mathbf{k})- \s_0(\mathbf{k})|^2,
\end{equation}
with $\s(\mathbf{k})$ from equation~\eqref{eq:structurefactor} and $Q$ a domain around the origin. 
Starting from an initial distribution for $\{\mathbf p_j\}$, the targeted patterns can then be obtained via gradient-based minimization methods with point coordinates as variables. Here, the Broyden–Fletcher–Goldfarb–Shanno (BFGS) minimization algorithm is considered. Efficient approaches to numerically solve this problem can be found in Refs.~\cite{Morse2023,Aaron2024}.

As in reference~\cite{salvalaglio2024persistent}, we use the following  parametrization for $\s_0$:
\begin{equation}\label{eq:formulaofstruc}
    \s_0(\mathbf{k})= 
    \begin{dcases}
        K^{-\alpha}(1-H)|\mathbf{k}|^{\alpha}+H \qquad |\mathbf{k}|< K\\
        \qquad \qquad \quad1 \qquad\qquad \quad \;\;\text{else}.
    \end{dcases}
\end{equation}
In this formula, $\alpha$ controls the power-law scaling of the structure factor for $|\mathbf{k}|\,$$<$$\,K$. Large $\alpha$ allows for approximating the so-called \textit{stealthy} HU character, i.e. $\s(\mathbf{k})\,$$\sim$$\, 0$ for $|\mathbf{k}|\,$$<$$\,K$. $K$ determines the wave numbers for which $\s$ deviates from $1$, i.e., it effectively sets the length scale above which correlations are imposed. For 2D stealthy HU configurations, an order-disorder transition is obtained for $K_{\rm c}\,$=$\,\sqrt{8\pi N}$ \cite{torquato2018hyperuniform}. 
$H$ controls the limit of the structure factor for $|\mathbf{k}|\rightarrow 0$.

The structure factor in Eq.~\eqref{eq:formulaofstruc} is chosen as it features a simple, straightforward parameterization while enforcing global HU characters and well-defined and comparable small-scale randomness. The latter follows from choosing $\s_0(\mathbf{k})=1$ for $|\mathbf{k}|\,$$>$$\,K$. Other forms for $\s_0(\mathbf{k})$ other than \eqref{eq:formulaofstruc} can be considered. In particular, several HU systems exhibit oscillations for $|\mathbf{k}|\,$$>$$\,K$ \cite{torquato2018hyperuniform}. Modeling these cases, however, requires introducing additional choices concerning the functional form and parameters while not introducing any qualitative novel aspect.
Another approach may consist of not enforcing any value of $\s_0(\mathbf{k})$ for $|\mathbf{k}|\,$$>$$\,K$. 

We recall that $H=0$ for an ideal HU pattern of infinite size, as dictated by equation~\eqref{eq:HUdef2}. For a finite size pattern, however, only the structure factor for the smallest appropriate $|\mathbf{k}|$ can be computed, namely $\min(|\mathbf{k}|)\,$=$\,k_{\rm min}$, which may deviate from $0$ even for $H=0$. An effective value for the limit of the structure factor for $|\mathbf{k}|\rightarrow 0$ is thus given by
\begin{equation}\label{eq:htilde}
\widetilde H = \frac{\s(k_{\rm min})}{\s(K)}.
\end{equation}
with the normalization used to generally account for eventual peaks in the structure factor at $K$ \cite{torquato2018hyperuniform}.
For the parametrization used in this work, 
$\widetilde H\,$=$\,\s(2\pi)$ as $k_{\rm min}\,$=$\,2\pi$ being the domain a unit square and $\s_0(K)\,$=$\,1$.
The quantity $\widetilde H$ is typically considered in realistic settings, e.g., experiments or numerical simulations, to practically define the HU character. The commonly adopted terminology refers to \textit{nearly HU} patterns for $H\leq10^{-2}$ and \textit{effective HU} patterns for $H\leq10^{-4}$ \cite{torquato2006hardspheres,kim2018imperfectionhyperuniformity}. In this article, for discussing geometrical features of HU arrangements, we will consider that a pattern is hyperuniform if $H\leq10^{-2}$. In \cite{salvalaglio2024persistent}, it was indeed shown that local arrangements in patterns of the size considered here are not or barely distinguishable for $H<10^{-2}$, motivating our choice. As detailed below, this condition will be practically sampled by means of the accessible $\widetilde H$ defined in equation~\eqref{eq:htilde}.

\begin{figure*}
    \centering
    \includegraphics[width=\linewidth]{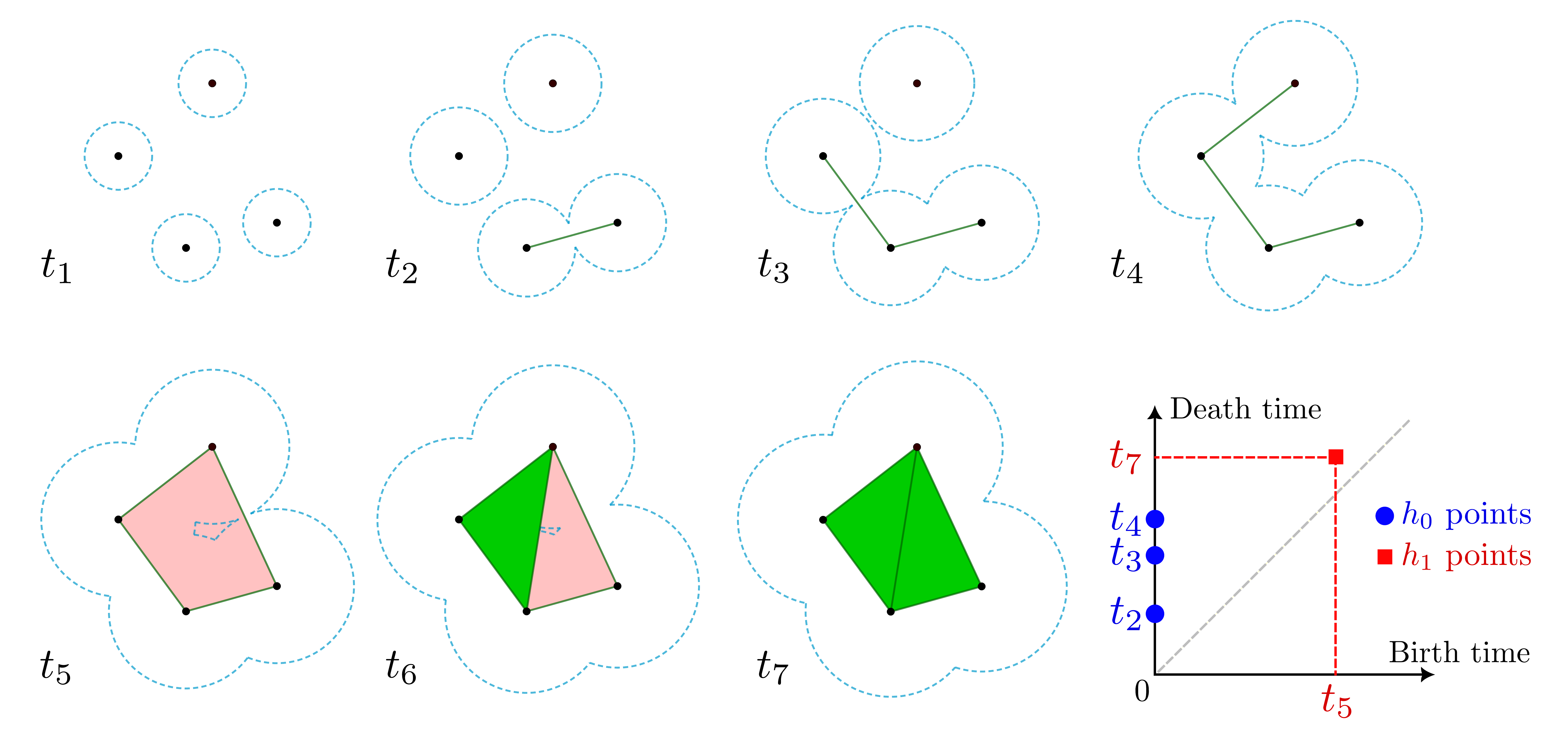}
    \vspace{-15pt}
    \caption{\label{fig:ph} Illustration of the \v{C}ech persistent homology and the persistence diagram for a point pattern of four points. 
    }
\end{figure*}

We study patterns of 780 points in a domain corresponding to the unit square with periodic boundary conditions. This specific number allows for having a good approximation of an ordered triangular lattice for large $K>K_{\rm c}\approx 45 \pi$ (with large $\alpha$ and $H$ almost~$0$) \cite{Uche2004,salvalaglio2024persistent}, thus ensuring proper sampling of a broad range of HU conditions. 
We chose $Q = [-50\pi, 50\pi]^2$ as the square region over which we minimize the objective function defined in equation~\eqref{eq:tominimize}. 
Using this configuration, we generated $1080$ different point patterns. More precisely, we generated $3$ samples for each possible combination of the three parameters $\alpha$, $H$ and $K$ with the following values:
\begin{equation}\begin{split}
    \alpha&\in \{0; 0.25; 0.5; 1; 2; 3; 5; 10; 20; 100 \},\\
    H&\in \{5\cdot10^{-4}; 10^{-3}; 10^{-2}; 0.1; 0.35; 1\}, \\
    K&\in\{28\pi; 34\pi; 40\pi; 45\pi; 51\pi; 56\pi\}.
    \label{eq:valuesofparam}
    \end{split}
\end{equation}
Any initial arrangement can be, in principle, considered for the minimization of $O(\{\mathbf{p}_j\})$, equation \eqref{eq:tominimize}.
An efficient approach consists of setting as initial configurations slightly (randomly) perturbed triangular lattices (different for every realization). This choice allows for faster convergence to the desired pattern compared to initial fully random arrangements. We note that the perturbation is needed to escape the unstable equilibrium represented by periodic lattices. Examples of generated patterns are given in figure~\ref{fig:points_Sk}. 

\section{Persistent homology}\label{sec:persitent-homology}

 Persistent homology is a technique used in topological data analysis, leveraging concepts from algebraic topology, to identify and characterize features of data that are persistent across multiple scales \cite{edelsbrunner2002topological,edelsbrunner2008persistent}. We refer to~\cite{edelsbrunner2022computational,otter2017roadmap,oudot2020inverse_persistence} and reference therein for general and comprehensive treatments of persistent homology while recalling here the main, practical concepts only for point patterns in two spatial dimensions (2D).

This technique typically considers families of simplicial complexes, i.e., collection of points, lines, triangles, tetrahedrons, and their higher-dimensional counterparts, which can be constructed from the considered data set. In particular, families of nested simplicial complexes $X_t$ parameterized by a scalar, positive parameter $t$, called \textit{time}, such that $X_s \subseteq X_t$ if $s \leq t$ are considered; see figure~\ref{fig:ph}. Topological features, e.g., the number of connected components (or 0-dimensional holes) or 1-dimensional holes, are then evaluated for $X_t$ upon varying $t$. Their variation allows for the characterization of the shape of the considered data. For instance, persistent features like a broad region without points within an otherwise uniformly distributed point pattern could be detected as a peculiar 1-dimensional hole, featured in different $X_t$ for a significantly large interval of $t$-values compared to other 1-dimensional holes.

 In practice, rules must be specified to construct the simplicial complexes $X_t$ while varying $t$, a procedure called \textit{filtration}. In this article, extending the preliminary analysis reported in reference~\cite{salvalaglio2024persistent}, two approaches are considered. The first one evaluates the persistent homology exploiting the so-called \v{C}ech complex, also equivalent to the alpha complex constructed from the finite cells of a Voronoi tessellation. The second approach exploits the so-called Vietoris-Rips complex \cite{oudot2020inverse_persistence}. We recall their definitions and further information in \ref{app:simplicial-complexes}. Both approaches define simplicial complexes from the set of circles with radius $t$ centered on the points in the considered data set; see figure~\ref{fig:ph} explicitly showing a filtration via \v{C}ech complex for a pattern comprising four points. 
 In particular, this figure shows seven different time values $t_n$ with $n=1,\dots,7$, realizing a nested family of seven simplicial complexes. The time at which a homological feature appears is called \textit{birth time} ($b$). When it disappears, it is called \textit{death time} ($d$). See, for instance, the four connected components (or 0-dimensional holes) at $t_1$ born at $t_0=0$ and the disappearance of one of them at $t_2$ as three connected components can be then identified. 

 For each $k$-dimensional hole ($k=0,1$ in 2D) in the simplicial complexes obtained by varying $t$, the birth time $b^{k}$ (appearance) and the death time $d^{k}$ (disappearance) are thus determined. 
 A typical representation of these quantities is the so-called \textit{persistence diagram}, which considers points $\mathbf{p}_j^k\,$=$\,(b^{k}_j,d^{k}_j)$ in the corresponding Cartesian representation with birth and death times as $x$ and $y$ coordinates, also illustrated in figure~\ref{fig:ph} (bottom right).  
 Note that $b^{0}_j\,$=$\,0$, as connected components (0-dimensional holes) have a birth time of $t\,$=$\,0$ by definition. The points $\mathbf{p}_j^k$ characterizing the persistent homology and entering persistence diagrams are hereafter called ${h}_k$ \textit{points} and their collection $h_k$-\textit{diagram} for brevity. 
 Further discussions and illustrations can be found in reference~\cite{pun2022machine_learning_persistence}. 

 Persistent homology and persistence diagrams relate to a few concepts well discussed in the context of hyperuniformity. Decorating the points with spheres, similar to filtration, is an approach often used to generate two-phase media (mimicking particles or voids) \cite{Torquato2002random,torquato2018hyperuniform} exploited to study HU phases. Also, the filtration itself can be related to the concept of spreadability in a two-phase medium, namely the diffusion from one phase into a second phase, which has been found to correlate with HU characters~\cite{torquato2021diffusion}. We remark that persistence diagrams contain information concerning the geometry of the analyzed pattern. Information on neighborhoods of all the points is retained well beyond two-body information. Moreover, they have a close relationship to the bounded hole property, which all stealthy HU systems and some non-stealthy HU have \cite{zhang2017can,wang2023equilibrium,klatt2022wave}.
 
\begin{figure}[t]
    \centering
    \includegraphics[width=\linewidth]{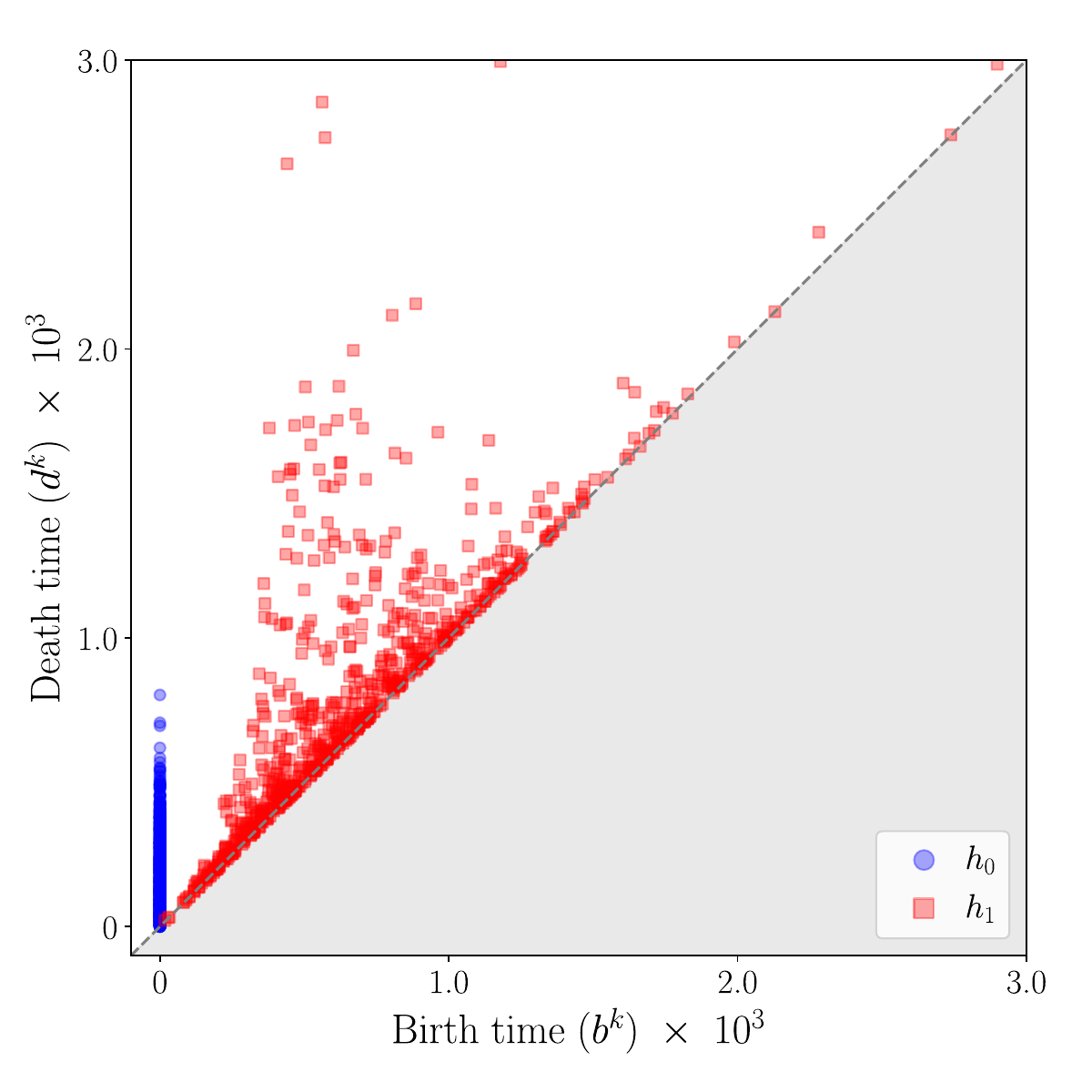}    
    \vspace{-15pt}
    \caption{\label{fig:pattern_and_diagram}  Persistence diagram for the point pattern reported in figure~\ref{fig:points_Sk} with  $\alpha\,$=$\,0.5$, $K\,$=$\,40\pi$ and $H\,$=$\,10^{-3}$. $h_0$ points $(b_j^0,d_j^0)$ (blue circles) denote the birth and death time of 0-dimensional holes (i.e., connected components). $h_1$ points $(b_j^1,d_j^1)$ (red squares) denote the birth and death time of 1-dimensional holes. }
\end{figure}

 For the evaluation of the persistence diagrams considering both Vietoris-Rips (\textit{RipsComplex}) or an alpha filtration (\textit{AlphaComplex}), we used the Python library \textit{gudhi}\cite{gudhi2015user_manual,rouvreau2015gudhi_alpha_complex,maria2016gudhi_rips_complex}.
 An example of (\v{C}ech) persistence diagram, featuring $h_0$ and $h_1$ points for the first pattern in figure~\ref{fig:points_Sk}, is reported in figure~\ref{fig:pattern_and_diagram}. 
 
 %
%

\section{Inferring hyperuniformity from persistence diagrams}\label{sec:machinelearning}

In this section, we show that the HU character of a point pattern can be inferred from 
a characterization of its local arrangements via persistent homology (introduced in section~\ref{sec:persitent-homology}). To this goal, we use standard machine learning (ML) methods \cite{pun2022machine_learning_persistence}, trained to encode the information contained in persistence diagrams. This is realized via selected features reconstructed from $h_0$ and $h_1$ diagram, presented below. We also outline the considered ML models and eventually report the results concerning the accuracy of the detection of hyperuniformity with the developed methods.

\subsection{The features}
\label{sec:mlmodel_features}

Persistent homology characterized via persistence diagrams as introduced in section~\ref{sec:persitent-homology} cannot be readily used for models based on ML. First, ML methods typically work with information or \textit{features} characterized by tens of values, while raw $h_k$ points are significantly more numerous: for $780$ points in the point patterns considered in this work, the persistence diagram would feature $\gtrsim 2000$ inputs ($779$ $h_0$ points and a number of $h_1$ points with the same order of magnitude); see the example in figure~\ref{fig:pattern_and_diagram}. Second, the number of $h_1$ points varies since the number of $1$-dimensional holes differs over different patterns: many ML models are not designed for receiving features characterized by a changing number of values. Therefore, generalized features that efficiently convert the data into a reduced and constant number of inputs are considered. We use three different ones. Two correspond to suitable features presented in reference~\cite{pun2022machine_learning_persistence}, while we introduced an additional one accounting for quantifying distances between the entire $h_1$ persistence diagrams at once. 

\subsubsection{Algebraic measures of the $h_k$ points.}
The \textit{first feature} takes algebraic measures of the coordinates of $h_k$ points. In particular, we consider these classes of measures:
\begin{itemize}
\item \textit{Adcock Algebraic} measures \cite{adcock2013algebraic}:
    \begin{equation}\label{eq:feat1}
    \begin{split}
        f_1^{k}=&\sum_j\frac{b_j^kl^k_j}{N_k}, \\  f_2^{k}=&\sum_j\frac{(\text{max}_n\{d^k_n\}-d_j^k)l^k_j}{N_k},\\ f_3^{k}=&\sum_j \frac{(b_j^k)^2(l^k_j)^4}{N_k},\\ f_4^{k}=&\sum_j\frac{(\text{max}_n\{d^k_n\}-d_j^k)^2(l^k_j)^4}{N_k}
          \end{split}
    \end{equation}
with $N_k$ the number of $h_k$ points $\mathbf{p}^k_j\,$=$\,(b_j^k,d_j^k)$ (see definition in section~\ref{sec:persitent-homology}), and $l^k_j\,$=$\,d^k_j-b^k_j$ their lifetime.
Since $k\,$=$\,0,1$, a feature array with $8$ inputs is obtained. 
\item \textit{Kali\v{s}nik algebraic} measures \cite{kalisnik2019algebraic}:
    \begin{equation}\label{eq:feat2}
    \begin{split}
        f_5^{k}=&\sum_j(l^k_j), \\f_6^{k}=&\text{max}_{i}(l^k_i), \\ f_7^{k}=&\text{max}_{i<j}(l^k_i+l^k_j),\\ f_8^{k}=&\text{max}_{i<j<m}(l^k_i+l^k_j+l^k_m),\\  f_9^{k}=&\text{max}_{i<j<m<n}(l^k_i+l^k_j+l^k_m+l^k_n), 
          \end{split}
    \end{equation}
with $l^k_j$ defined as above. Since $k\,$=$\,0,1$ a feature array with $10$ inputs is obtained. 
\end{itemize}

\begin{figure*}[t]
     \centering
     \includegraphics[width=\linewidth]{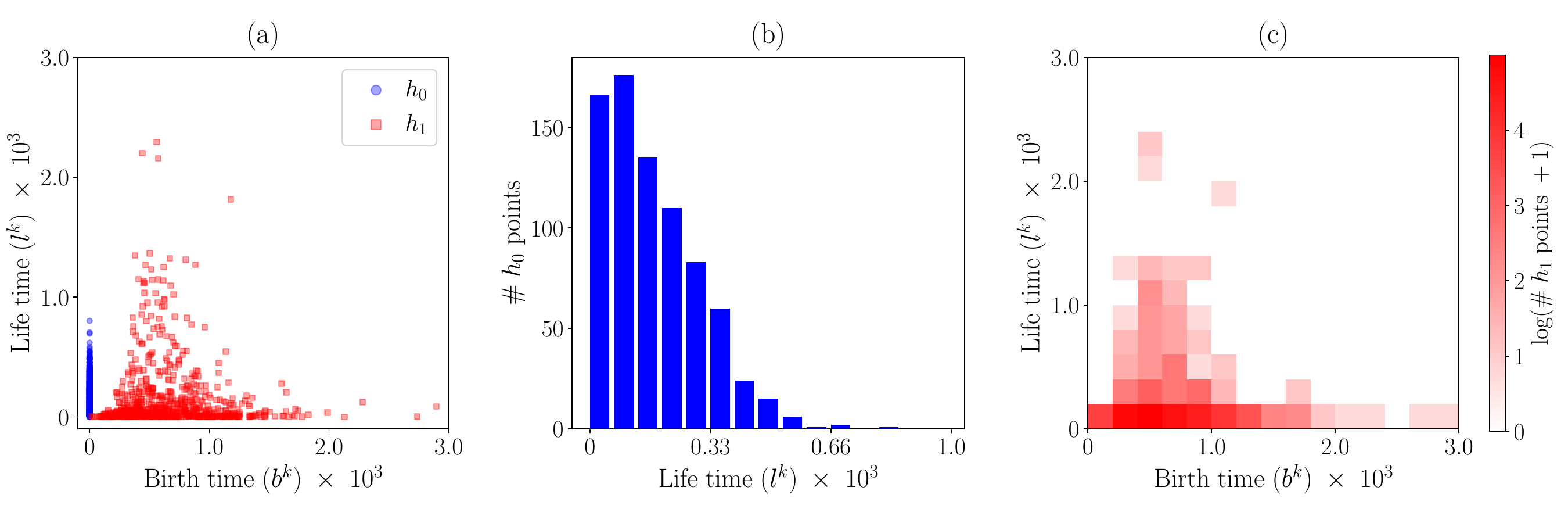}
     \vspace{-15pt}
     \caption{\label{fig:binning} Binning of the persistence diagram in figure~\ref{fig:pattern_and_diagram}. 
     (a) Modified persistence diagram neglecting the $b > d$ region by replacing the death times ($d_j^k$) with the life times ($l_j^k=d_j^k-b_j^k$), i.e. applying the mapping $(b_j^k,d_j^k)\mapsto (b_j^k, l_j^k)$. (b) Histogram of $h_0$ points in panel (a) for a suitable size of the (uniform) bins, here for $n=15$, with $n$ the number of bins. (c) A color map presenting the binning of $h_1$, here for $n=15$ with $n$ the binning of the axis (see further details in the text). The color indicates the number of $h_1$ points in each bin.}
 \end{figure*}

 \subsubsection{Binning of the persistence diagram.}
The \textit{second feature} results from the binning of the persistence diagram. A given observation window for the persistence diagram is chosen and sampled by a certain number of bins. Since the $h_k$ points cannot appear under the diagonal $b=d$ by construction, we do map them under $(b_j^k,d_j^k)\mapsto (b_j^k, d_j^k-b_j^k)\,$=$\,(b_j^k, l_j^k)$ to optimize the use of the observation window; see figure~\ref{fig:binning}(a). An undersampling of the $h_k$ points is then considered by counting their number over the bins. Owing to their equal birth time $b^0_j\,$=$\,0$, the binning of $h_0$ points can be described with an array of scalar values and thus represented as a histogram; see figure~\ref{fig:binning}(b). The binning of $h_1$ points results in a matrix and can be represented by a (discrete) heat map; see figure~\ref{fig:binning}(c).

The evaluation of this feature is performed as follows: for the Vietoris-Rips persistent homology, the observation interval of the $h_0$ points is $[0, 0.05]$, and the observation window of the $h_1$ points is $[0, 0.09]\times [0, 0.075]$. For the \v{C}ech persistent homology, they are respectively of $[0, 0.0005]$ and $[0, 0.002]\times[0, 0.0015]$. The difference in the size of the observation window comes from the associated different range of birth and death times as obtained with the implementation of the two homologies in \textit{gudhi}\cite{rouvreau2015gudhi_alpha_complex,maria2016gudhi_rips_complex}. Importantly, the resolution of the binning method has to be specified. For a number $n$, we can divide the observation interval of the $h_0$ points in $n$ bins and the observation window of the $h_1$ points in $n\times n$ bins, resulting in a feature having $n^2+n$ inputs. We used here $n=9$, $15$ and $23$.

   \subsubsection{Wasserstein distance between persistence diagrams.} \label{sec:mlmodel_features_wd}
The \textit{third feature} is based on the Wasserstein distance that quantifies the distance between two sets of points $A$ and $B$, particularly useful in various fields such as statistics, ML, and optimal transport theory \cite{villani2009optimal,panaretos2019statistical}. We consider the $q$-th Wasserstein distance
 \begin{equation}
 \begin{split}
 \label{eq:wasserdef}
 W_q(A,B)=\min_{U,\gamma} \bigg(&\sum_{\mathbf{a}\in U}\delta(\mathbf{a},\gamma(\mathbf{a}))^q+\sum_{\mathbf{a}\in A\setminus U}\delta(\mathbf{a}, \Delta(\mathbf{a}))^q
 \\ 
 &+\sum_{\mathbf{b}\in B\setminus \gamma(U)}\delta(\mathbf{b}, \Delta(\mathbf{b}))^q\bigg)^{1/q}
 \end{split}
 \end{equation}
 with $U\subseteq A$, $\gamma:U\rightarrow B$ injective, $\delta$ a distance function for pairs of points, and $\Delta$ the common projection on the diagonal $x\,$=$\,y$. Here, $\gamma$ is a partial matching between $A$ and $B$, a map from some elements of the former to some of the latter. Note that using the partial matching $\gamma$, this definition copes with $A$ and $B$ having a different number of elements. It evaluates the distance (according to the definition of $\delta$) of the matched pairs of points, but also by the distance to the diagonal of the unmatched points. Thus, $W_q(A,B)$ corresponds to the minimal distance among all these partial matches and measures the cost of transforming $A$ into $B$. 
 
 The Wasserstein-distance-based feature associated with a pattern is then built by measuring the distance $W_p$ of the corresponding $h_1$ diagram from a set of reference ones. As references, we consider here the $h_1$ diagrams of HU patterns (generated as in section~\ref{sec:hu}) with the following parameters:
\begin{equation}
\begin{split}
    &\alpha_{\rm R}\in [0.5, 3, 20], \\  &H_{\rm R}\in[0.0001, 0.05, 0.35], \\ &K_{\rm R} \in [28\pi, 40\pi, 51\pi],
    \label{valuesofref}
\end{split}
\end{equation}
thus corresponding to 27 qualitatively different (reference) structure factors. 
The Wasserstein-distance-based feature then corresponds to an array of 27 values $[W]_i$ with $i\in \{1, \dots ,27\}$. An average over 3 independent realizations for each combination of the parameters in \eqref{valuesofref} has also been considered to reduce possible spurious effects due to the randomness underlying the generation process. We verified that this relatively small number is enough to avoid such effects for the applications considered here. 
The calculation of the Wasserstein distance has been performed using the function \textit{wasserstein\_distance} of the python library \textit{gudhi} \cite{gudhi2015user_manual}. We used its default implementation corresponding to  $\delta(\mathbf{a},\mathbf{b})\,$=$\,L^{\infty}(\mathbf{a},\mathbf{b})\,$=$\,\max_i|a_i-b_i|$ and $q\,$=$\,1$.

\begin{table*}
    \caption{\label{tab:result2} Accuracy of the ML models in detecting hyperuniformity exploiting different features. VR: Vietoris-Rips, \v{C}: \v{C}ech}
    \vspace{10pt}

\centering
\footnotesize
\begin{tabular}{@{}l||lllllll}
         \hline
         \diagbox{\bf Features}{\bf ML Model} & \makecell{linear SVM} & \makecell{SVM\\ with RBF} & \makecell{Pruned Tree} & \makecell{Random \\Forest} & \makecell{Boosted Tree} & \makecell{Simple\\ Neural \\Network} & \makecell{Deep\\ Neural \\Network} \\
         \hline    
         \hline    
         \vspace{-4pt}
         \\
         Adcock Algebraic & \makecell{VR: 72\%\\ \v{C}: 72\% } & \makecell{VR: 92\%\\ \v{C}: 90\% } & \makecell{VR: 92\%\\ \v{C}: 72\% } & \makecell{VR: 92\%\\ \v{C}: 72\% } & \makecell{VR: 92\%\\ \v{C}: 72\% } & \makecell{VR: 91\%\\ \v{C}: 89\% } & \makecell{VR: 91\%\\ \v{C}: 89\% } \\[11pt]
         Kali\v{s}nik Algebraic & \makecell{VR: 84\%\\ \v{C}: 72\% } & \makecell{VR: 84\%\\ \v{C}: 83\% } & \makecell{VR: 90\%\\ \v{C}: 92\% } & \makecell{VR: 92\%\\ \v{C}: 92\% } & \makecell{VR: 92\%\\ \v{C}: 91\% } & \makecell{VR: 90\%\\ \v{C}: 91\% } & \makecell{VR: 91\%\\ \v{C}: 90\% }\\[11pt]
         Binning $n=9$ & \makecell{VR: 92\%\\ \v{C}: 94\% }& \makecell{VR: 92\%\\ \v{C}: 91\% }& \makecell{VR: 91\%\\ \v{C}: 93\% }& \makecell{VR: 95\%\\ \v{C}: 95\% }& \makecell{VR: 94\%\\ \v{C}: 96\% }& \makecell{VR: 93\%\\ \v{C}: 94\% }& \makecell{VR: 93\%\\ \v{C}: 94\% }\\[11pt]
         Binning $n=15$ &\makecell{VR: 94\%\\ \v{C}: 93\% } &\makecell{VR: 92\%\\ \v{C}: 92\% } &\makecell{VR: 91\%\\ \v{C}: 92\% } &\makecell{VR: 94\%\\ \v{C}: 94\% } &\makecell{VR: 94\%\\ \v{C}: 95\% } &\makecell{VR: 94\%\\ \v{C}: 93\% } &\makecell{VR: 95\%\\ \v{C}: 94\% } \\[11pt]
         Binning $n=23$ &\makecell{VR: 93\%\\ \v{C}: 94\% } &\makecell{VR: 93\%\\ \v{C}: 92\% } &\makecell{VR: 92\%\\ \v{C}: 91\% } &\makecell{VR: 94\%\\ \v{C}: 94\% } &\makecell{VR: 94\%\\ \v{C}: 94\% } &\makecell{VR: 94\%\\ \v{C}: 94\% } &\makecell{VR: 94\%\\ \v{C}: 94\% }\\[11pt]
         Wasserstein distance& \makecell{VR: 93\%\\ \v{C}: 84\% } & \makecell{VR: 92\%\\ \v{C}: 92\% } & \makecell{VR: 92\%\\ \v{C}: 93\% } & \makecell{VR: 94\%\\ \v{C}: 94\% } & \makecell{VR: 94\%\\ \v{C}: 94\% } & \makecell{VR: 95\%\\ \v{C}: 94\% } & \makecell{VR: 95\%\\ \v{C}: 94\% }\\[11pt]
         \hline
    \end{tabular}
\end{table*}
\normalsize

\subsection{The ML models}
\label{sec:mlmodel_descriptions}

To assess if hyperuniformity can be inferred from persistent homology via the features illustrated above, we explore several ML models reported in the review \cite{pun2022machine_learning_persistence}: linear and radial basis function (RBF) support vector machine (SVM), pruned tree, random forest, boosted tree, simple and deep neural networks. 

In brief, for the SVM methods, we used the \textit{Support Vector Classification (SVC)} function of \textit{scikit-learn}. We used both a linear kernel, searching for a linear separation between classes of data, and the RBF kernel, using the kernel method to project the data into a higher dimensional space and search for a linear separation. From a different perspective, other models called decision trees search for sets of conditions that can differentiate classes of data. They can be implemented in \textit{scikit-learn} with the function \textit{DecisionTreeClassifier}. These models have a strong tendency to overfit. Therefore, it is necessary to prune them by removing the less effective decision nodes (\textit{pruned tree}). In \textit{scikit-learn}, this is achieved by tuning the parameter \textit{ccp\_alpha}, whose best value is set via tests on the training set. Other methods can also be considered: the \textit{random forest} (\textit{RandomForestClassifier}) generates several decision trees with different subsets of the data to minimize overfitting. In this case, we chose to generate $100$ decision trees. Furthermore, the choice of the trees can be improved by using some linear regression, giving rise to \textit{boosted trees}, implemented by \textit{GradientBoostingClassifier}. For them, we set $200$ as the number of classifiers. Finally, given their potential, we also used neural networks. These are implemented in \textit{scikit-learn} by the function \textit{MLPClassifier}, and we designed two specific networks: the first one possesses only one \textit{simple} hidden layer with $128$ neurons, the second one is \textit{deep} and possesses three hidden layers, having $128$, $128$ and $256$ neurons respectively. As these can all be considered standard methods, we refer to \cite{pun2022machine_learning_persistence} and references therein for additional information. A more general and complete introduction to ML models can be found in \cite{Bishop2006}.


The training of the ML models was performed as follows: first, we extracted the features detailed in section~\ref{sec:mlmodel_features} from the persistence diagrams of patterns with an ideal structure factor generated as described in section~\ref{sec:hu}. Then, for each possible pair of features and ML models among the ones illustrated above, we have trained the model to detect hyperuniformity from the corresponding feature. To this goal, we labeled a point pattern as HU or non-HU via a binary variable according to a condition on $\widetilde H$, i.e. the effective smaller value of the structure factor at the smallest sampled wavenumber; see equation~\eqref{eq:htilde}. We consider a pattern HU if $\widetilde H<0.011$ approximating the nearly-HU condition $ H \leq 0.01$ (as $\widetilde H$ typically overestimates $H$ for small $\alpha$, while approaching its value for large $\alpha$ and large systems only; see figure~\ref{fig:points_Sk}). 
The persistent homology leveraging Vietoris-Rips and the \v{C}ech complexes are considered, with independent training and testing phases. We used 5-fold cross-validation, i.e., we divided evenly and randomly the $1080$ patterns into 5 sets, and we trained the ML models $5$ times, training with $4$ of the sets and testing with the last one. This also enables us to compute a mean result as well as the standard deviation.

\begin{figure*}
     \centering
     \includegraphics[width=\textwidth]{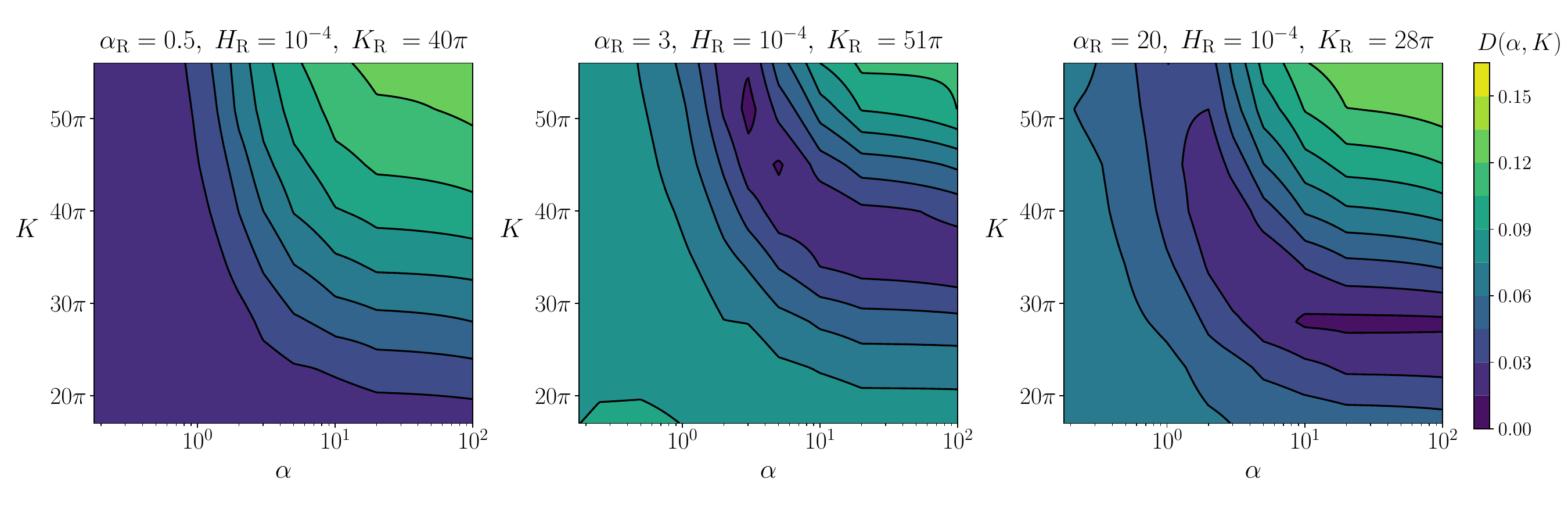}
     \vspace{-15pt}
     \caption{\label{fig:distancemaps} Examples of three distance maps $D_i(\alpha,K)$ used to determine effective structure factor parameters of a given pattern. Parameters of the corresponding reference arrangements ($\alpha_{\rm R}$, $H_{\rm R}$, $K_{\rm R}$) are reported above the panels.
     }
 \end{figure*}

\subsection{Results}
\label{sec:mlmodel_results}

The accuracy in the detection of the HU character in point patterns by varying features and ML models in the training is summarized in Table~\ref{tab:result2}. The standard deviations range between $0.5\%$ and $3\%$ and are all reported in Table~\ref{tab:result2_SD} in \ref{app:standard deviation}. We can first observe that the hyperuniformity is overall well detectable from the local geometry of arrangements as described in persistence diagrams, up to an accuracy of $(96 \pm  1.5)\% $.
Such accuracy meets values typically considered very successful for ML algorithms combined with persistent homology \cite{pun2022machine_learning_persistence}. These results thus unveil the existence of a very significant correlation between persistent homology, i.e., the local geometry of the arrangements in the patterns, and hyperuniformity. Thus, it shows that inverse criteria with respect to the systematic dependence of persistent homology on structure factor parametrization disclosed in reference~\cite{salvalaglio2024persistent} can be devised.

 More precisely, we observe that the best results are obtained with Boosted Tree and Neural Networks ML models combined with the binning and the Wasserstein-distance-based features. This supports the usage of the latter, newly proposed here, for the analysis of HU arrangements and suggests its exploration for other kinds of patterns and applications as well. Both the different simplicial complexes considered to evaluate the persistent homology (Vietoris-Rips and \v{C}ech, see section~\ref{sec:persitent-homology}) deliver similar performances over most of the considered models, except in the case of Algebraic features. Being too dependent on the specific considered persistent homology, they should then be discarded for general applications of the method. It is worth mentioning that the linear SVM provides very good results with the binning feature up to $(94\pm 2)\%$ with $n\,$=$\,9$. Besides its simplicity, this points to a relatively robust linear separation in the feature space of HU and non-HU arrangements, which may be exploited for classification algorithms and generalizations based on classical (linear) regression methods.

Although we focused here on a classification problem, features based on persistence diagrams and suitable ML can be applied to regression problems as well, provided suitable data sets are available and/or can be generated. In \ref{app:estimationH}, we show that such a framework allows for estimating the value of the parameter $H$ directly, confirming the possibility of inferring detailed characteristics of hyperuniformity from persistence diagrams.

\section{Estimation of effective structure factor parameters from persistence diagram}\label{sec:compute_parameters}

In the previous section, we have shown that the hyperuniformity in terms of criteria based on $\widetilde H$ can be inferred from persistent homology. This has been done without discerning between different kinds of hyperuniformity, dictated in our framework by parameters like $\alpha$ and $K$ in equation~\eqref{eq:formulaofstruc}. In this section, we proceed further with our analysis and show that such parameters, and thus specific HU characters, can also be determined.

We focus here on HU patterns only, by setting $H\,$=$\,10^{-3}$, and their characterization via $h_1$ \v{C}ech diagrams. 
We first compute the Wasserstein-distance-based feature introduced in section~\ref{sec:mlmodel_features_wd} for different values of $\alpha$ and $K$. Wasserstein distance maps $D_i(\alpha,K)$ with $i\in \{1, \dots ,27\}$ are considered upon linear interpolation between neighbors in the $(\alpha, K)$ space (with an additional average of two realizations of the patterns associated to $\alpha$ and $K$). Selected examples of $D_i(\alpha,K)$ are reported in figure~\ref{fig:distancemaps}.
The effective ($  \alpha$, $  K$) parameters of an independent pattern, given its Wasserstein-distance-based feature $[W]_i$, can then be determined by finding the ones at which $D_i(  \alpha,   K)\approx  [W]_i$.
This corresponds to the minimization of the following objective function
\begin{equation}\label{eq:obj2}
    \mathcal{O}'(  \alpha,    K)=\sum_{i=1}^{27}\left(D_i( \alpha,   K) - [W]_i \right)^2,
\end{equation}
performed here by exploiting the conjugate gradient method, repeated over different initial guesses for $  \alpha,    K$ to cope with the existence of local minima.


\begin{figure*}
     \centering
     \includegraphics[width=0.85\textwidth]{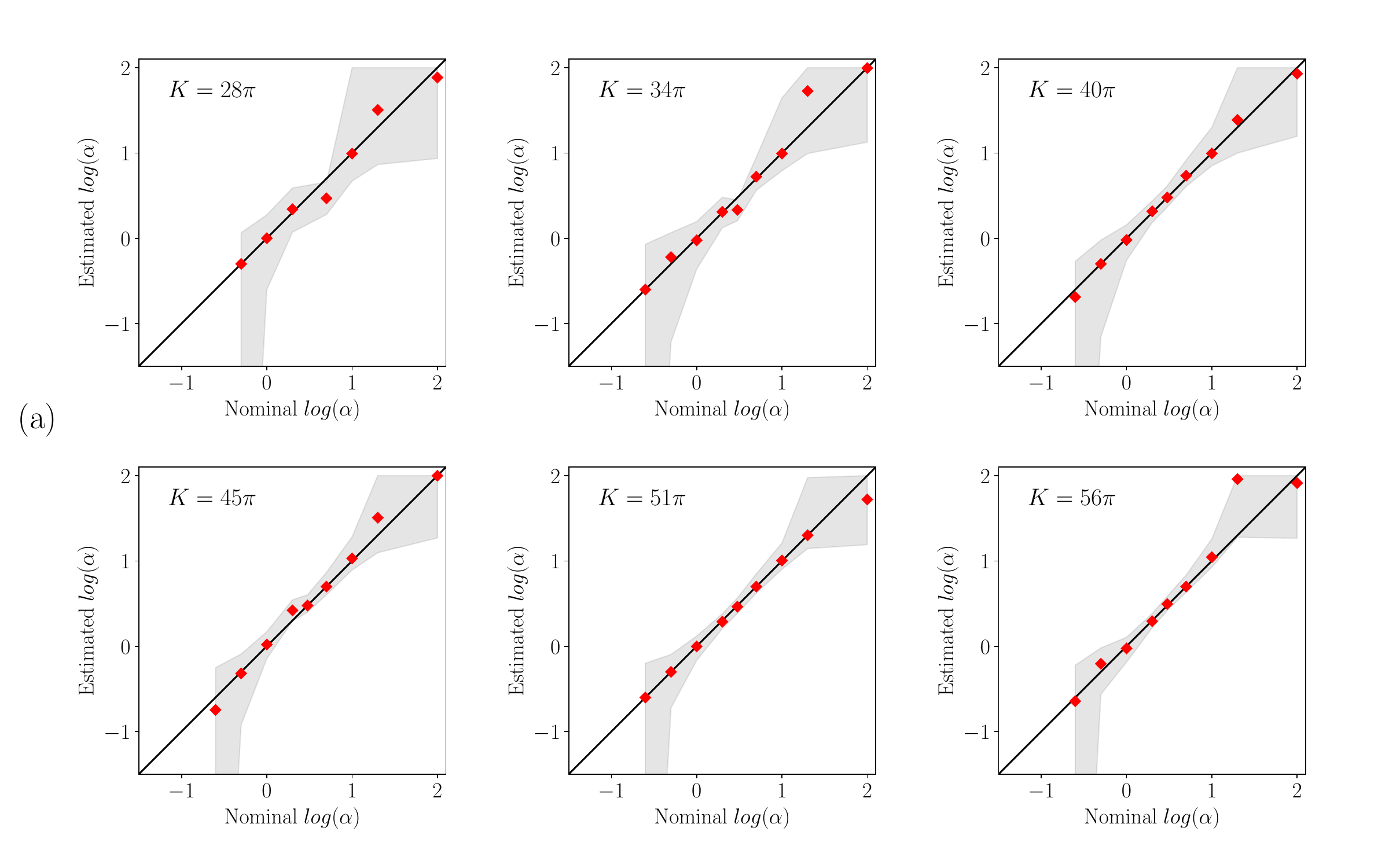}
     \includegraphics[width=0.85\textwidth]{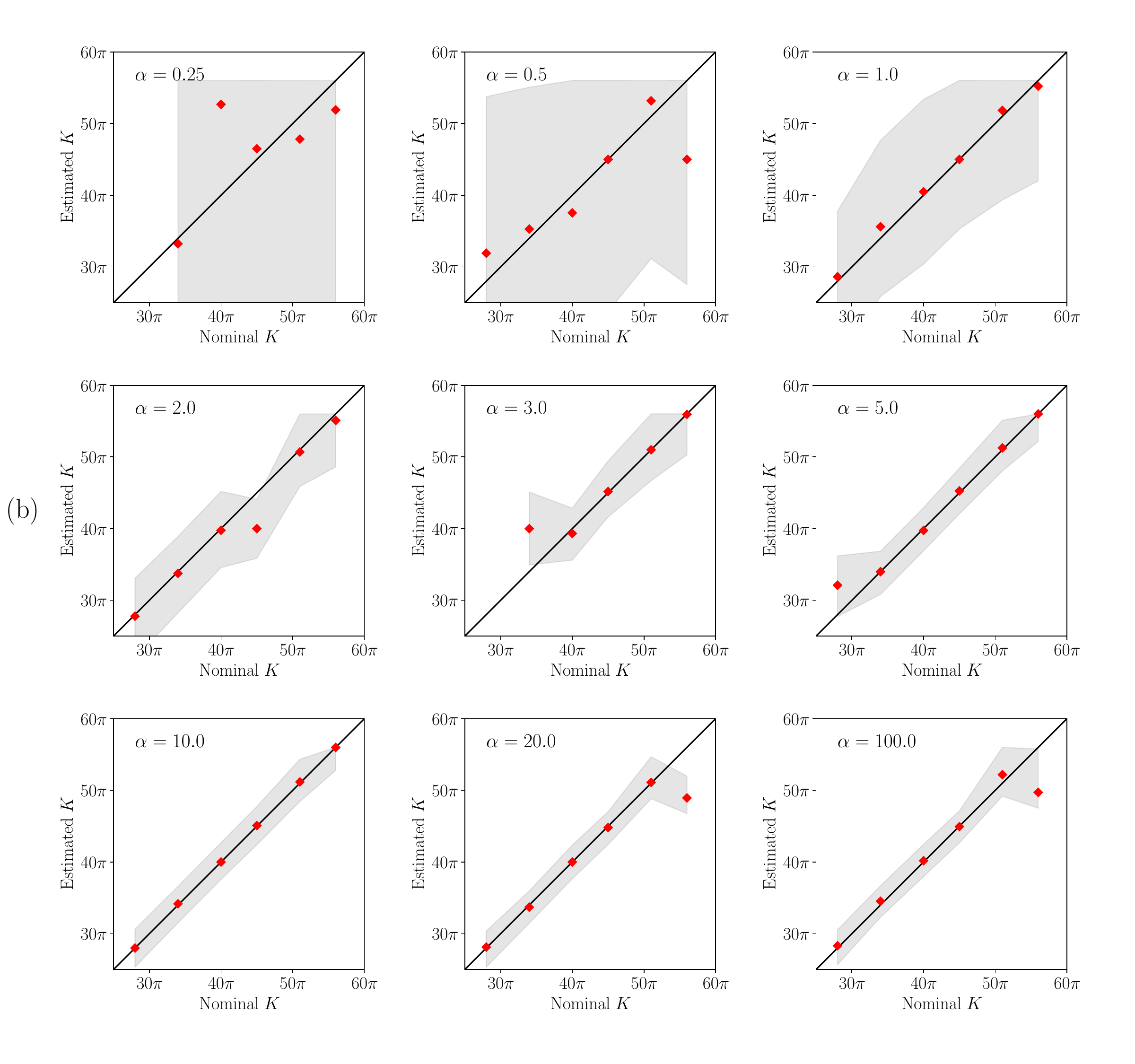}
     \vspace{-15pt}
     \caption{\label{fig:resultdistance} Results of the parameter estimation from the persistence diagram. (a) Estimations of $\alpha$ (red squares) for different nominal $K$ values. (b) Estimations of $K$ (red squares) for different nominal $\alpha$ values. For each plot, a confidence interval (gray region) indicates the range around the estimated parameter for which $\mathcal{O}'(  \alpha,    K)$ differs less than $\epsilon\,$=$\,0.002$ from the minimum. Even though the results of $\alpha$ and $K$ are presented separately for a matter of readability, they are estimated together.}
 \end{figure*}

An overview of the estimation of effective $  \alpha$ and $  K$, always simultaneously determined via the minimization of $\mathcal{O}'$, is reported in figure~\ref{fig:resultdistance}. Therein, estimated parameters (red dots) and a confidence interval (gray area) are reported. The latter is determined as the range of the parameters for which $\mathcal{O}'(  \alpha,    K)$ differs from less than a given (small) threshold $\epsilon$ from the found minimal value.

For a broad range of parameters, the estimated $  \alpha$ and $  K$ values correspond well with the nominal values used to generate the patterns. Focusing first on the estimation of $\alpha$ for different nominal values of $K$, figure~\ref{fig:resultdistance}(a), we see that for intermediate nominal values $ 1 < \alpha < 10$, the prediction is very accurate with a narrow confidence interval. 
For the smallest sampled nominal $\alpha$, we observe that the estimation of $\alpha$ is accompanied by a broadening of the confidence interval. This points to less characteristic differences in the persistence diagrams of the corresponding patterns, expected due to a more significant contribution of randomness (see figure~\ref{fig:points_Sk} for $\alpha\,$=$\,0.5$). Enough information is still retained to obtain estimation of $\alpha$ very close to nominal ones.
Large nominal $\alpha$ values are also associated with a broadening of the confidence interval and, overall, a relatively accurate estimation of the nominal parameter when considering values within the confidence interval. This result may be ascribed to very similar topological arrangements realized for small variations of the parameters. Indeed, for large $K$ values (above $K_{\rm c}$, see section~\ref{sec:hu}), nearly ordered patterns -- namely triangular arrangements with defects -- are obtained (see figure~\ref{fig:points_Sk}(a) for $K\,$=$\,51\pi$ and $\rm \alpha=100$). As such, minimal differences between corresponding persistence diagrams are realized when varying the structure factor parameters. This is also realized in the stealthy HU patterns obtained for smaller $K$ values.

The trends discussed above are further made evident in figure~\ref{fig:resultdistance}(b), showing the estimation of $  K$ for given nominal $\alpha$ values. Significantly large confidence intervals and deviation from nominal parameters are obtained for $\alpha\,$=$\,0.25$ up to $\alpha\,$=$\,0.5$ over the whole range of nominal $K$ values. Progressively more accurate predictions are obtained when increasing $\alpha$ for a broad range of $K$ values. Note that, however, no broadening of the confidence interval is obtained for large $\alpha$ when estimating $K$, pointing out that variation of this parameter produces a more significant effect in ordered and stealthy HU than $\alpha$, which is then easier to detect.

\begin{figure}[h]
     \centering
     \includegraphics[width=\linewidth]{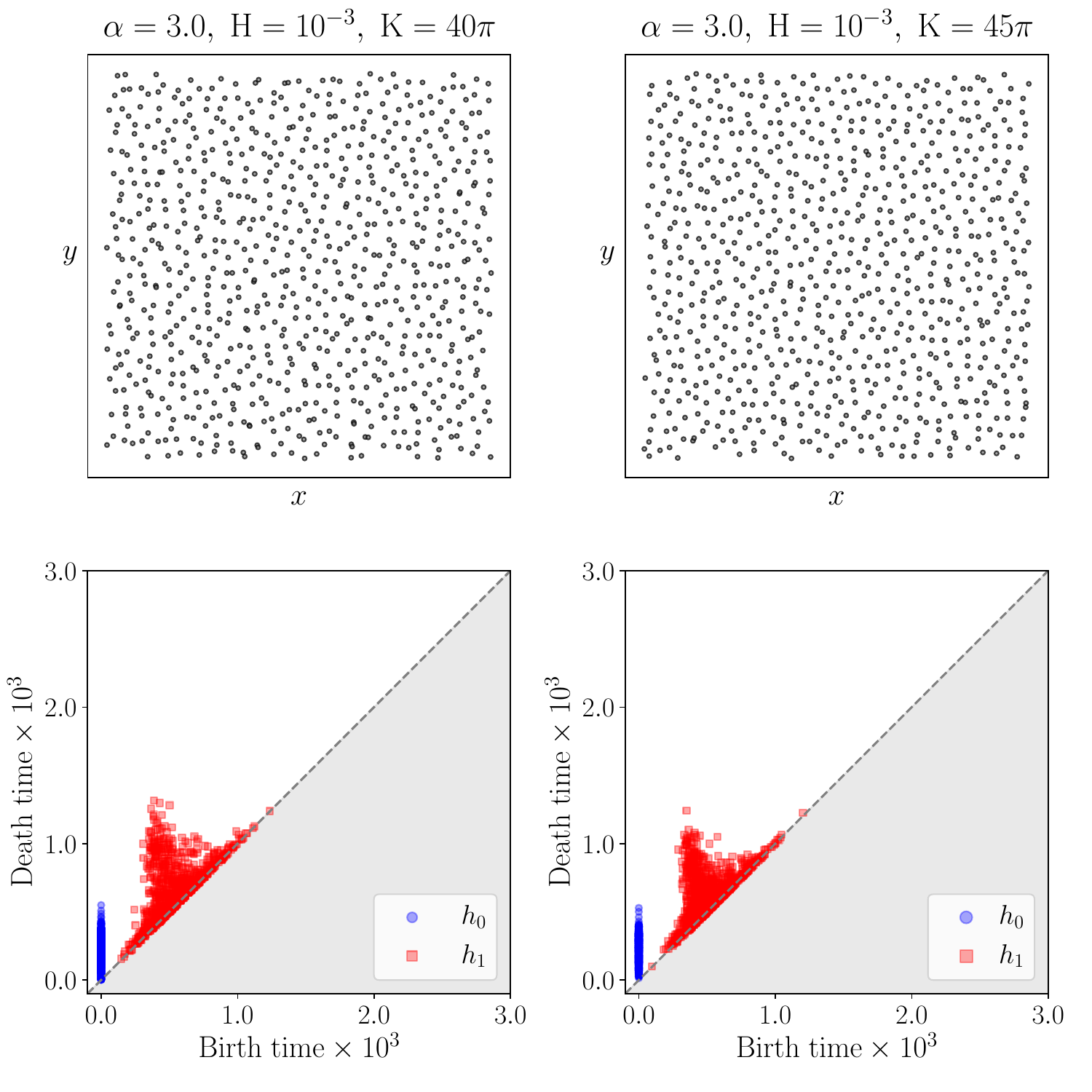}
     \vspace{-15pt}
     \caption{\label{fig:diffk} 
     Two point patterns (first row), with nominal structure factor parameters reported above the panels, and (second row) their corresponding \v{C}ech
 persistence diagrams. Although they have a similar appearance, they are distinguished by the approach discussed in section~\ref{sec:compute_parameters}; see figure~\ref{fig:resultdistance}(b), $\alpha=3.0$ panel. 
 }
 \end{figure}

From these results, we may conclude that details set via global constraints in HU patterns can also be devised by looking at features of the local arrangements, with minor limitations for ordered and highly disordered systems only.
Importantly, in the results of figure~\ref{fig:resultdistance}, the nominal values are almost always contained in the confidence interval. We remark that this interval, although arbitrary, points to the region with very similar arrangements.
For example, in figure~\ref{fig:resultdistance}(b) for $\alpha\,$=$\,3$, the confidence interval has a width of $\approx$$\,5\pi$ in $K$. Patterns generated with $\alpha\,$=$\,3$, $H\,$=$\,10^{-3}$, and the two different nominal values $K=40\pi$
and $K\,$=$\,45\pi$, together with their persistence diagrams, are shown in figure~\ref{fig:diffk} and are clearly not easily distinguishable. Still, although featuring extremely similar geometrical arrangements and persistence diagrams, information concealed in the latter is enough to distinguish them and correctly estimate the parameter of their structure factor (as shown in figure~\ref{fig:resultdistance}(b)).  
Such a small but detectable difference obtained by varying $K$ by $\pm 5\pi$ is further used in the following to assess the accuracy in generating a pattern approximating a given persistence diagram.

We also remark that although patterns with different structure factors may exhibit very similar arrangements, as can be directly seen by extended regions with distances approaching 0 in the panels of figure~\ref{fig:distancemaps}, exploitable differences remain as shown here with the correct estimation of the specific parameters used for the generation of the patterns in the first place, especially when considering comparisons to many reference pattern at ones (as done via the objective function \eqref{eq:obj2}). 

\section{Detecting boundaries between disordered arrangements}
\label{sec:boundaries}

A scenario that can be handled efficiently by persistent homology is the coexistence of different point patterns. It resembles a situation very common in ordered systems like (poly)crystals, where domains with a different order or orientations (grains) are separated by defected regions (grain boundaries) \cite{ma2017random}. Inhomogeneities emerge at the boundaries of disordered systems, which we show below are captured by the information contained in persistence diagrams.


 \begin{figure*}[h]
     \centering
     \includegraphics[width=\linewidth]{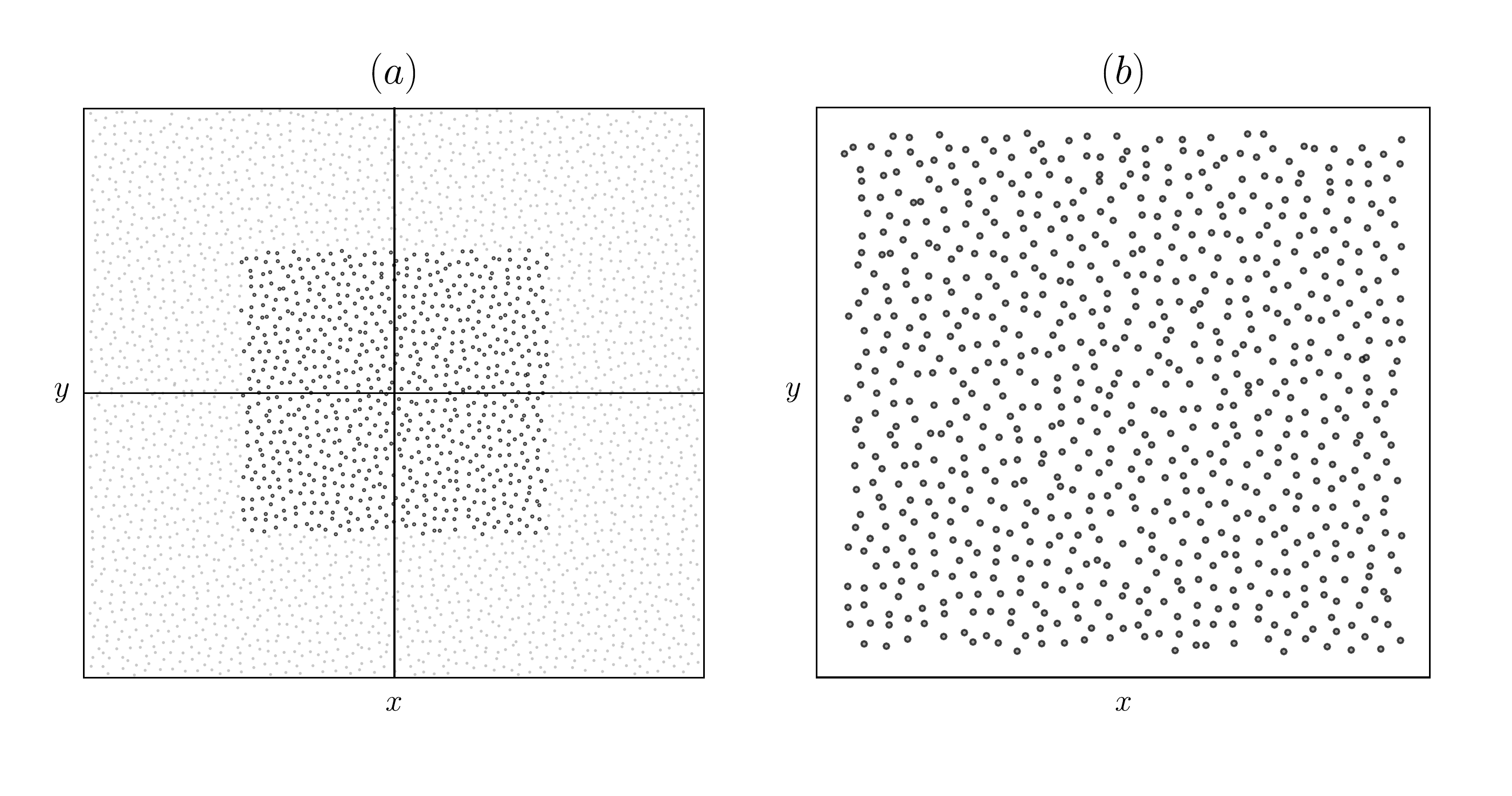}
     \vspace{-15pt}
     \caption{\label{fig:stitch_method} 
     Four-grains pattern. (a) construction method: four different homogeneous patterns sharing the same structure factor (here $\alpha=3.0$, $H=10^{-3}$ and $K=51\pi$) are stitched together, and a central square containing $780$ points (dark grey) is selected. (b) Resulting four-grains pattern (dark grey in panel a). 
 }
 \end{figure*} 


A new arrangement exhibiting boundaries between point patterns featuring ideal HU characters is generated. Four different realizations of ideal HU patterns (780 points) with the same structure factor are juxtaposed into a large square pattern with $4\times780$ points. To have a comparable persistence diagram between this new ``four-grain" setting and the ideal HU patterns (``one-grain"), a square region centered in the middle and containing $780$ points is selected. This construction is illustrated in figure~\ref{fig:stitch_method}. We focus here on patterns featuring a marked HU character with $\alpha\geq 3$, $H=10^{-3}$ and $K\geq 45\pi$.

\begin{figure*}[h]
     \centering
     \includegraphics[width=\linewidth]{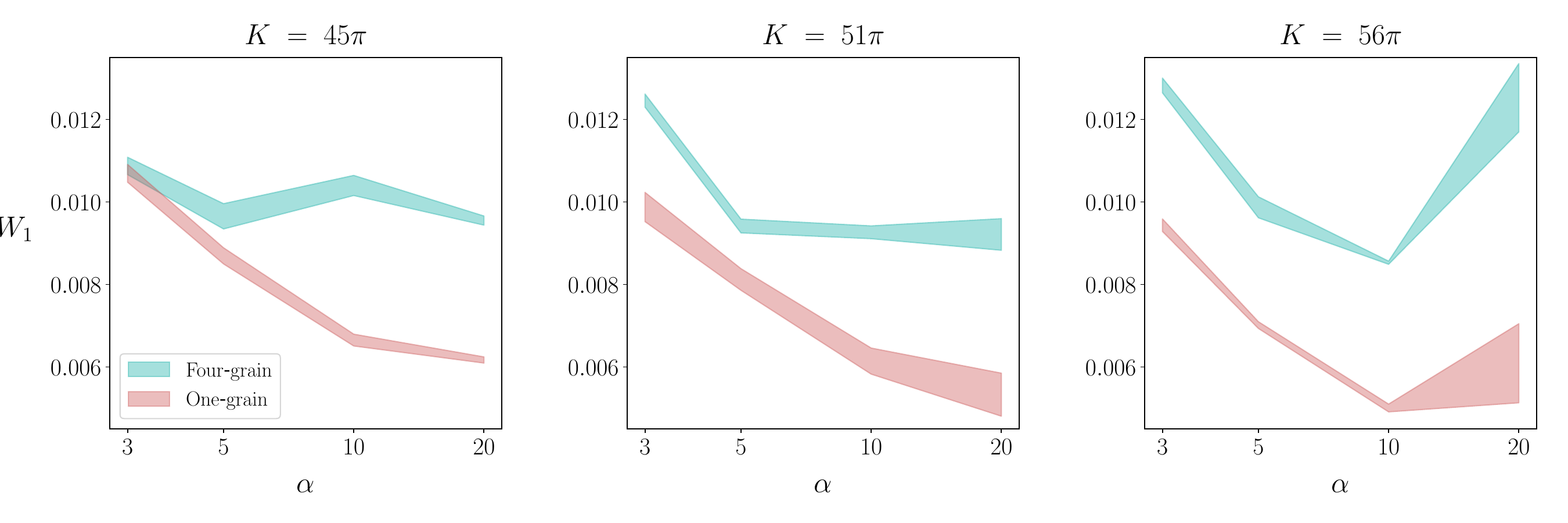}
     \vspace{-15pt}
     \caption{\label{fig:stitch_results} 
     Wasserstein distance between the persistence diagrams of one-grain from both other one- and four-grain patterns (blue-green and red regions, respectively). Colored regions are centered on the mean Wasserstein distance with a width equal to the standard deviation over several independent evaluations. 
 }
 \end{figure*}

To quantify the difference between the persistence diagrams of four- and one-grain patterns, we use the Wasserstein distance as introduced in equation \eqref{eq:wasserdef}. More precisely, we evaluated this distance 12 times with different four-grain and one-grain realizations, obtaining a mean value as well as a standard deviation. These results can then be compared to the typical distance between two one-grain patterns (as in previous sections) extracted from 6 different evaluations. The obtained standard deviation intervals are summarized in figure~\ref{fig:stitch_results}. They clearly show that the presence of boundaries is encoded in the persistence diagrams by a significant increase in their (Wasserstein) distance from those of ideal, one-grain patterns. As an illustration, the case with $\alpha=3.0$ and $K=51\pi$ presented in figure \ref{fig:stitch_method} induces a difference of around $20$\% of the distances. In other words, the presence of boundaries between ``grains" is effectively captured by the persistent homology. 

This result further confirms the suitability of the considered method for the study of local geometry beyond homogeneous arrangements. Moreover, by focusing on subsets of inhomogeneous patterns, methods can be envisaged to detect where the boundary regions are located, if any. Cases involving overlap of patterns of empty space between patterns would lead to even more significant features in the persistence diagrams and can be treated similarly to the case explicitly discussed in this section. 


\section{Exploring the construction of patterns from (HU) persistence diagrams}
\label{sec:reconstruction}

 In this section, we explore the construction of a point set reproducing a given persistence diagram. In this way, we may then obtain a point pattern encoding targeted topological features, focusing here on the case of the persistence diagrams realized in HU arrangements without constraints on global properties, e.g., the domain of definition of the arrangement, boundary conditions, or its structure factor. This represents the first step toward designing configurations that enable the independent examination of their unique geometries, separate from the global HU characteristics.
 
 Inverting persistence diagrams to obtain the originating point patterns is currently a subject of active research. It is known that several point patterns share the same $h_1$ points (see, e.g., \cite{kurlin2022identical_persistence}). On the other hand, it was shown in reference~\cite{gameiro2016continuity_persistence} that the variation of $h_k$ diagrams is locally smooth regarding variations of point-pattern coordinates and that specific persistence diagrams may be enforced via a minimization procedure starting from an initial condition close enough to the targeted one, including having the same number of $h_1$ points. This latter constraint is, however, too restrictive in our case. Moreover, these concepts have been applied to simple topological features, namely to persistence diagrams with just a few $h_1$ points \cite{gameiro2016continuity_persistence}, so largely deviating by the data set we consider (see, e.g., figure~\ref{fig:pattern_and_diagram}). 

We thus introduce a procedure to construct arrangements whose geometry approximates a given persistence diagram. For the sake of simplicity, we restrict the discussion to the \v{C}ech persistent homology, recalling that it usually contains more information than the Vietoris-Rips (see \ref{app:simplicial-complexes}), and consider the topological information encoded in $h_1$ diagrams. All the details are reported in \ref{app:pattern-from-homology} while we summarize here the main steps. 

\begin{figure*}
     \centering
     \includegraphics[width=\linewidth]{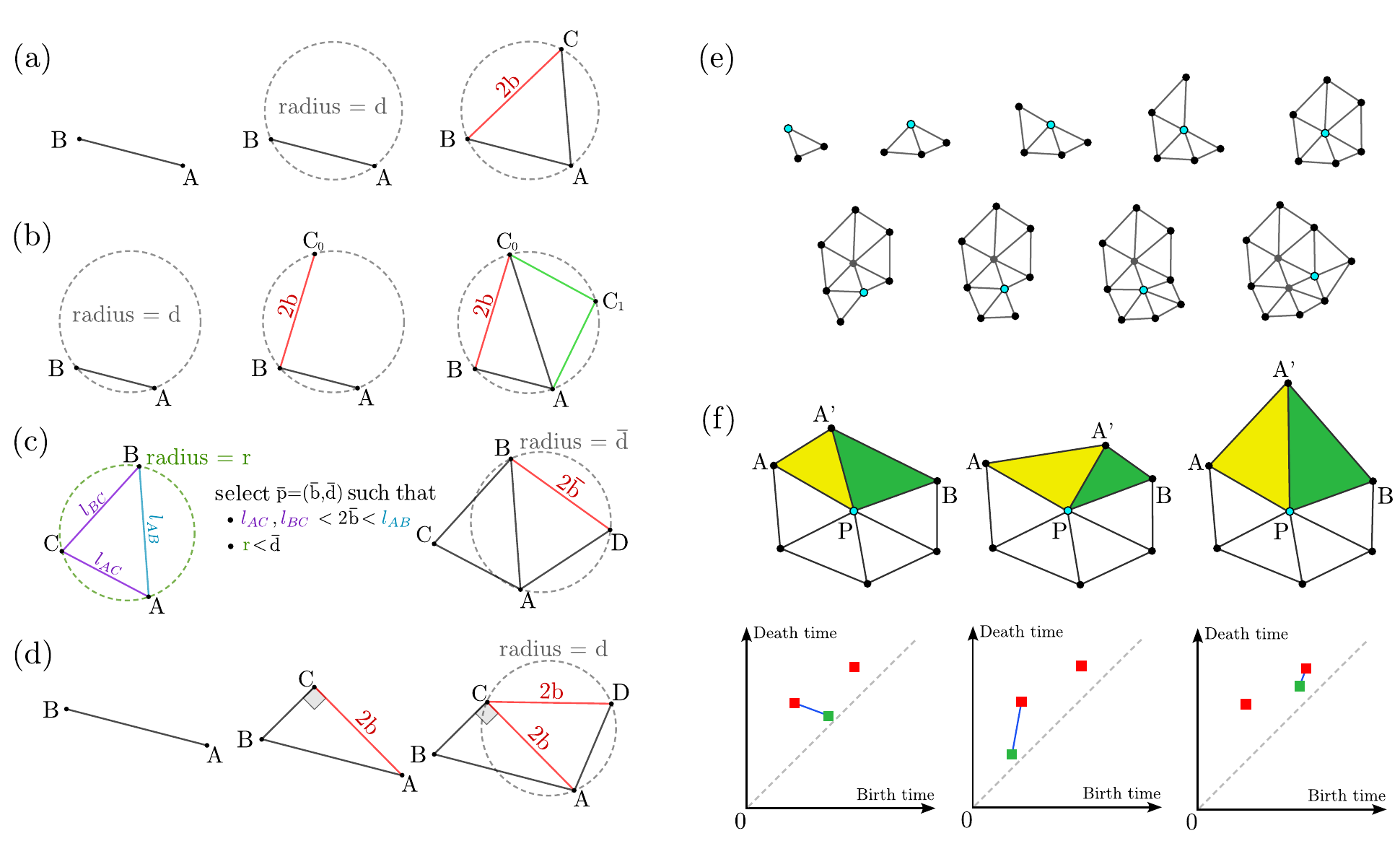}
     \vspace{-15pt}
     \caption{\label{fig:constructions} Illustration of the different basic constructions of triangles from $h_1$ points. $A$ is the point selected to apply the general construction algorithms (the \textit{marked point} in the algorithms in \ref{app:pattern-from-homology}), $AB$ is a given edge with length $l_{\rm AB}$ and $p=(b,d)$ a given $h_1$ point. While (a) is the standard construction, (b) is applied when the edge $AB$ is too short for a correct application of the first one. If $AB$ is too large, (c) can be used if a $h_1$ point $\bar p$ satisfying the indicated conditions can be found. If not, (d) is applied.
     All these geometrical constructions are detailed in \ref{app:pattern-from-homology}.
     (e) An example of the recursive construction of triangles around a \textit{marked point}, indicated in light blue.
     (f) Illustration of the closing procedure. By varying the position of $A'$, we construct the AA'P (yellow) triangle to reproduce one of the targeted $h_1$ points, $p_1$. This implies the definition of the additional PA'B (green) triangle that is described by another $h_1$ point, $p_2$ (green squares in the persistence diagrams). While $p_1$ is in the targeted persistence diagram, $p_2$ is generally not. Different positions of $A'$ and thus realizations of $p_1$ and $p_2$ are then inspected, and the one for which $p_2$ is the closest to one of the targeted $h_1$ points (red squares in the persistence diagrams) is selected.
     }
 \end{figure*}

 \begin{figure*}
     \centering
     \includegraphics[width=0.9\linewidth]{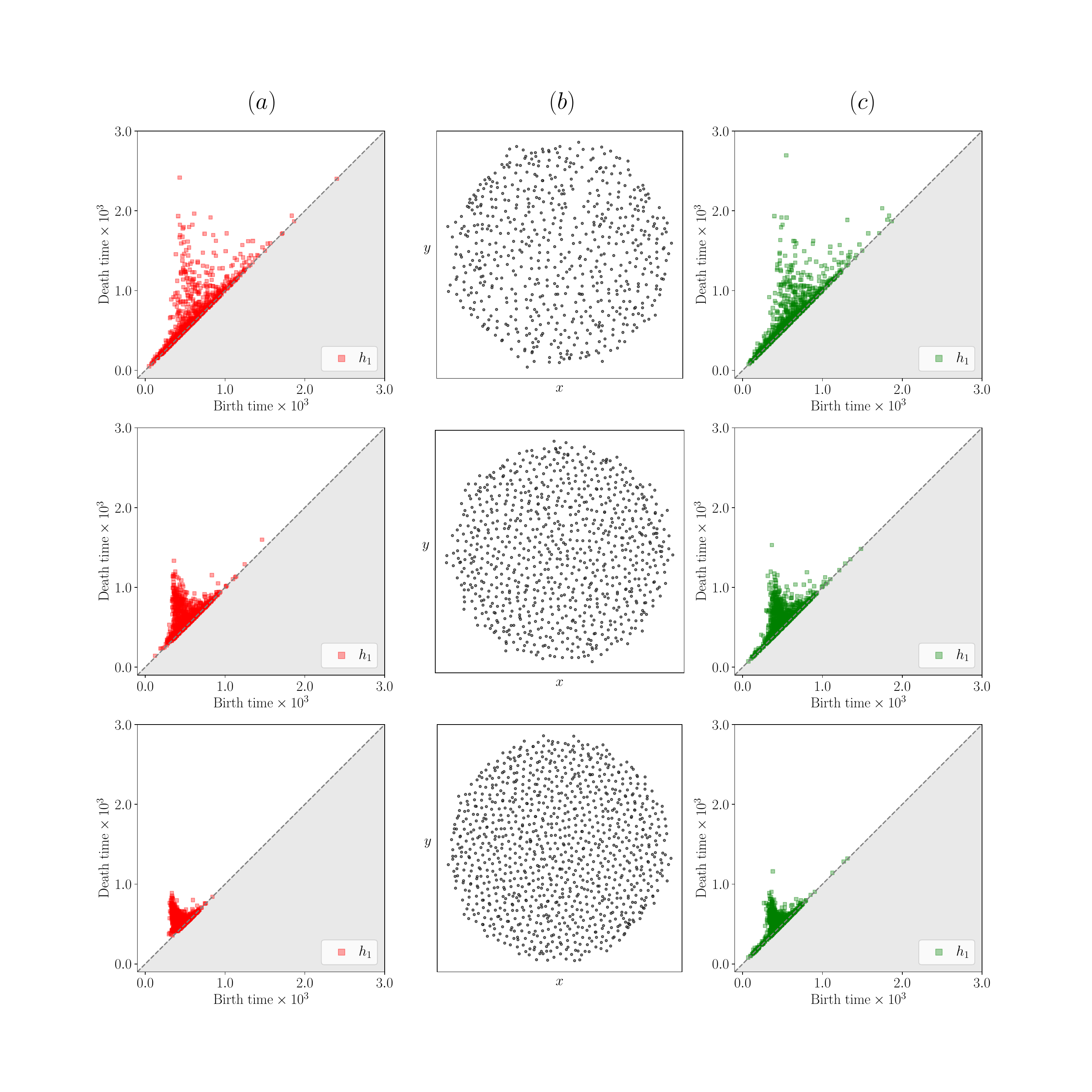}
     \vspace{-15pt}
     \caption{\label{fig:produced_patterns} Representative results demonstrating the generation of patterns with persistence diagrams that closely approximate the target ones. Three examples corresponding to the three rows are shown. (a) Original persistence diagrams --obtained here from ideal HU arrangements-- to approximate. (b) Constructed point patterns via the procedure outlined in section~\ref{sec:proc} from the persistence diagrams in panel (a). (c) Persistence diagrams of the point patterns in panel (b), which approximate the ones in panel (a). The parameters of the patterns from which the persistence diagrams in panel (a) originate are from top to bottom: (i) $\alpha\,$=$\,1.0$, $H\,$=$\,5\cdot 10^{-4}$, $K\,$=$\,34$, (ii) $\alpha\,$=$\,3.0$, $H\,$=$\,5\cdot 10^{-4}$, $K\,$=$\,45$ and (iii) $\alpha\,$=$\,5.0$, $H\,$=$\,5\cdot 10^{-4}$ and $K\,$=$\,56$.}
 \end{figure*}

\subsection{Construction algorithm}
\label{sec:proc}

In the proposed algorithm, we begin from a triangle that is compatible with having an arbitrary selected $h_1$ point (see, e.g., the single $h_1$ point in figure~\ref{fig:ph}), and then construct recursively triangles juxtaposed to the previously constructed ones. This implies considering an already existing edge, a segment $AB$, and we construct triangles from it corresponding to the available $h_1$ points in the persistence diagram until all of them are represented. A crucial aspect is that sliver triangles --triangles with one angle larger than $90^\circ$-- are avoided since they do not produce any $h_1$ point and may modify the topological features of the surrounding structure they are juxtaposed to. Some other configurations with non-sliver triangles may affect the topological features as well, and the devised algorithm tends to restrict or avoid their appearance. 

In brief, once a $h_1$ point $(b,d)$ is arbitrarily selected, the construction of one triangle $ABC$ inscribed in a circle with radius $d$ as illustrated in figure~\ref{fig:constructions}(a) is chosen if it is non-sliver while having the largest edge with a length equal to $2b$. Otherwise, two cases are possible: i) $AB$ is too short, i.e., one of the newly constructed edges would have a length greater than $2b$ or $ABC$ is sliver, in which case the construction of more triangles is considered as shown in figure~\ref{fig:constructions}(b); ii) $AB$ has itself a length strictly greater than $2b$. In this case, we select another $h_1$ point from the persistence diagram for which the first mentioned construction can be applied; see figure~\ref{fig:constructions}(c). If no $h_1$ point satisfying these conditions is found, the construction illustrated in figure~\ref{fig:constructions}(d) is considered by choosing another arbitrary $h_1$ point. In contrast to the previous ones, this latter method may affect the topological features of the already constructed arrangement.

Using these methods, triangles are then constructed around a given point according to expected topological features; see, for instance, figure~\ref{fig:constructions}(e). The last iteration leading the point surrounded by triangles requires a closing procedure: different constructions are sampled, and the one associated with $h_1$ points closely resembling points in the targeted persistence diagram is selected. See figure \ref{fig:constructions}(f) for an illustration.
Once the point is surrounded by triangles, the same process is repeated on a peripheral point, i.e., a point not surrounded by triangles, where the peripheral point the closest to $(0,0)$ is chosen. 
The whole procedure is reported in detail in \ref{app:pattern-from-homology}: in Algorithm \ref{alg:general}, the general method is described, while the construction of triangles and the closing procedures therein are summarized in the sub-Algorithms \ref{alg:construction} and \ref{alg:closing}, respectively.

\begin{figure*}
    \centering
    \includegraphics[width=\linewidth]{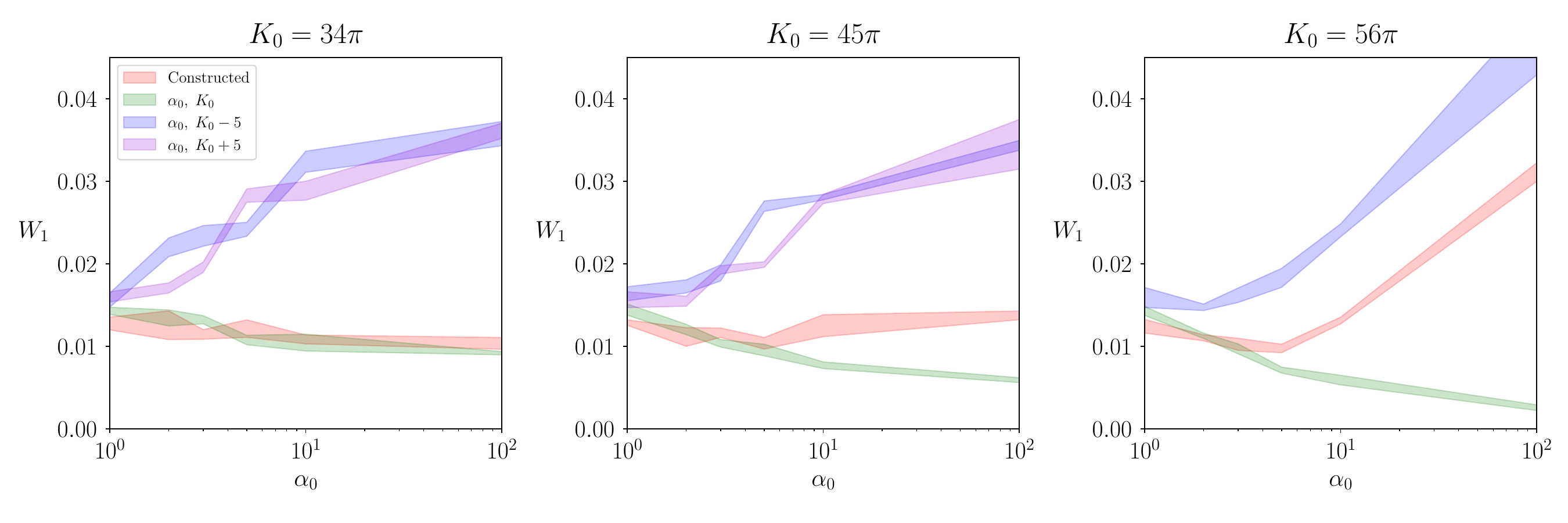}
    \vspace{-15pt}
    \caption{\label{fig:wasserstein_orginal_and_constructed} Quantitative analysis of the generation of patterns targeting a given persistence diagram obtained from a HU pattern with structure factor parameters $(\alpha_0,K_0,H\,$=$\,5 \cdot 10^{-4})$. The Wasserstein distance ($W_1$) between the targeted persistence diagram and those computed for i) another independent realization of the pattern (green) ii) the constructed one according to the method outlined in section~\ref{sec:proc} (red), iii) and iv) pattern with parameters $(\alpha_0,K_0\pm 5,H\,$=$\,5 \cdot 10^{-4})$ are reported. The curves are shown as shaded regions centered on the mean and with a width equal to the standard deviation over five independent evaluations of the distances above. The three plots are obtained for selected values of $K_0$ while varying $\alpha \in [10^0,10^2]$. Note that case $K= (56+5)\pi$ is not plotted as the ordered state is already reached for $K= 56\pi$. }
\end{figure*}

\subsection{Results}

We apply the algorithm outlined above to $18$ different persistence diagrams, all computed from different HU patterns with $H\,$=$\,5 \cdot 10^{-4}$. Some examples of targeted persistence diagrams and generated patterns are shown in figure \ref{fig:produced_patterns}(a) and \ref{fig:produced_patterns}(b), respectively. The persistence diagram computed from these newly generated patterns, shown in \ref{fig:produced_patterns}(c), qualitatively well matches the original one, suggesting that the goal to approximate the targeted topological features has been reached. Some deviations between the persistence diagrams in \ref{fig:produced_patterns}(a) and \ref{fig:produced_patterns}(c) are still observed, such as ``tails" along the diagonal. Note that without constraints on the region over which the points are generated and proceeding with local information only, the proposed construction method delivers a point pattern that has a different global shape (a circle vs the square domain used to generate the ideal HU patterns as detailed in section~\ref{sec:hu}). Consequently, and due to the locality of the construction, no boundary conditions are to be expected. Note that a consequence is that the structure factor of the constructed pattern inherently deviates from the one of the pattern from which the targeted persistence diagram is computed.  

To quantitatively assess the accuracy of the algorithm, we compute the Wasserstein distance ($W_1$) of the original $h_1$ diagram (e.g. \ref{fig:produced_patterns}(a), with $\alpha_0$ and $K_0$ the parameters of the point pattern which generated it) to: i) the $h_1$ diagram of the reconstructed pattern (as, e.g., \ref{fig:produced_patterns}(c)), ii) the persistence $h_1$ diagram of a different ideal pattern generated with $\alpha\,$=$\,\alpha_0$, $K\,$=$\,K_0$, iii) the persistence $h_1$ diagrams of patterns generated with $\alpha\,$=$\,\alpha_0$, $K\,$=$\,K_0\pm 5$ (mimicking small variations as in figure~\ref{fig:diffk}). The results summarized in figure~\ref{fig:wasserstein_orginal_and_constructed} show that the algorithm performs very well quantitatively. Persistence $h_1$ diagrams of the original patterns have a $W_1$ distance from the diagrams of another ideally generated pattern (green) similar to the constructed one (red). The distances from the persistence $h_1$ diagrams for patterns with $\alpha\,$=$\,\alpha_0$, $K\,$=$\,K_0\pm 5$ (implying only tiny geometrical changes as explained in section \ref{sec:compute_parameters}) always largely exceed these values. This is obtained for all the analyzed cases. Still, consistent with the results of the previous section, the algorithm is found to perform very well for disordered HU patterns, while it shows some larger deviation from ideality for large $\alpha$, particularly when leading to ordered systems at large $K$ (see the $K\,$=$\,56\pi$ panel in figure~\ref{fig:wasserstein_orginal_and_constructed}).

We may conclude that the proposed approach can reconstruct a pattern with given topological features as described by the persistent homology (here shown for the persistence $h_1$ diagrams in particular) with high accuracy. Therefore, this method can be exploited to reproduce the local arrangements imposed by global characteristics, as explicitly shown here for the ones realized in HU patterns.

\section{Discussion and Conclusions}
\label{sec:conclusions}

We investigated whether local arrangements realized in point patterns can be used to deduce HU or non-HU characters. It is shown that inverse relations between hyperuniformity and persistent homology can be inverted in an approximated fashion. This is achieved by leveraging persistent homology and data-driven techniques, and we illustrate how such a task can be performed. The results show that the information contained in persistence diagrams allows for characterizing and using local (geometrical and topological) features and, in general, can be used to detect arrangements analogous to the ones realized in HU patterns. We envision this latter aspect to be crucial for characterizing systems with the emergence of hyperuniformity in confined regions. 

We first showed that a binary classification ``HU vs. non-HU", determined according to a representative threshold on $\widetilde{H}$, can be devised from persistence diagrams. Diverse ML algorithms, trained to detect hyperuniformity associated with different derived local topological features, succeeded in recognizing this property with high accuracy, pointing to the robustness of such evidence. This result also allows for stating that hyperuniformity and persistent homology are closely related, with direct correlations first assessed in reference~\cite{salvalaglio2024persistent} and inverse ones newly unveiled here.

Quantitative comparisons between persistence diagrams allow for the estimation of global properties, as shown in terms of the parameters entering the (ideal) structure factor of the point pattern from which persistence diagrams are computed.
We remark that the HU character can be, in general, easily computed from equation~\eqref{eq:structurefactor} by looking at the power law decay for small $\mathbf{k}$ as well as $\widetilde{H}$, in infinite or periodic systems.
In this regard, the proposed framework can be used as a detection method, but the direct calculation of the structure factor (or the number variance) would be more accurate and efficient. The estimation of the parameters discussed here, however, shows that information contained in distributions of local structures can be used for quantitative estimations. 

 More interesting is the detection of traits of hyperuniformity in non-periodic, finite-size systems, e.g., where the underlying physics may enforce a HU character (e.g., the physical systems in Ref.~\cite{salvalaglio2020hyperuniform,Backofen24,Zheng2024}) while finite domain size may disrupt the ideal HU behavior. The latter is indeed rigorously defined for infinitely extended domains only (and mimicked in periodic domains). For finite-size systems, either extrapolation or improved statistical estimations \cite{hawat2023estimating} should be considered, although defect perturbations and finite size effects induce additional phenomenology (see, e.g., \cite{chen2021topological,Tsabedze_2022,PuigDEF,MilagrosBesana_2024}), which can be hardly fully captured by global measures only.
In these cases, estimating effective structure factor parametrization based on local structures proposed here may thus represent a practical method to characterize the system comprehensively, accounting for additional effects from small- to intermediate scales. In addition, applications of ML methods and reference persistence diagrams can be envisaged to screen large datasets efficiently.

We recall that our analysis builds on the model structure factor introduced in Eq.~\eqref{eq:formulaofstruc}. Besides introducing a specific, tunable HU (or non-HU) character at large length scales, it enforces the same randomness for all the patterns at small scales. We expect that the considered framework based on data analysis and advanced statistical methods, namely persistent homology and ML methods, would benefit from any additional information for small scales ($|\mathbf{k}|\,$$>$$\,K$), such as other functional forms or even an unprescribed structure factor --thus adapting in an uncontrolled manner to global order-- for the corresponding range of wave numbers. As the analysis of the distribution of local structures proposed here is done by constraining the pattern to feature ideal randomness at short scales, we expect that the result reported here represents a lower bound in terms of accuracy for other strongly correlated HU systems.

Given that we show that HU arrangements entail peculiar local arrangements and/or geometrical properties, we envision a detailed inspection of the dependence of physical properties on the latter, for instance, in the transient phase of transport phenomena in porous media \cite{TORQUATO2020103565} or diffusion \cite{torquato2021diffusion}. It will be essential to generate a pattern with targeted topological features independently from global constraints to study how related effects can be decoupled, thus augmenting the design space. It is in this context that the method we proposed to generate (point) patterns given the topological features encoded in persistence diagrams is particularly interesting, besides addressing a theoretical, fundamental aspect. We remark that the developed method may find application well beyond HU patterns and in the more general context of persistent homology.

A natural direction for future research consists of extending the proposed analysis and the study of persistent homology to scalar and vectorial fields exhibiting hyperuniformity. Examples are two-phase media obtained by decorating point patterns \cite{Torquato2002random} with circles or other shapes as well as considering parametrization for prototypical HU fields such as spinodal patterns \cite{ma2017random}. This also includes inspecting different parametrizations of the structure factor to study specific physical systems. Inverse design methods for generating HU arrangements entailing specific properties will also be explored. In this regard, we have shown that features based on persistent homology correspond well to HU characters. With this work, we have then provided crucial information for their exploitation as descriptors for inverse and data-driven inverse design concepts.

\section*{Acknowledgements}
We acknowledge useful discussions with Axel Voigt, Markus Kästner, and Ivo Sbalzarini. This work was supported by the German Research Foundation (DFG) within the Research Training Group GRK 2868 D$^3$ - project number 493401063. Computing resources have been provided by the Center for Information Services and High-Performance Computing (ZIH), the \href{www.nhr-verein.de/unsere-partner}{NHR Center} of TU Dresden.

\section*{Data Availability Statement}
The data that support the findings of this study will be made openly available in suitable repositories in the final version.

\addcontentsline{toc}{section}{ORCID iDs}
\section*{ORCID iDs}

Abel H. G. Milor \orcidlink{0009-0001-3890-0608} \href{https://orcid.org/0009-0001-3890-0608}{https://orcid.org/0009-0001-3890-0608}\\
Marco Salvalaglio \orcidlink{0000-0002-4217-0951} \href{https://orcid.org/0000-0002-4217-0951}{https://orcid.org/0000-0002-4217-0951}

\appendix

\renewcommand{\thetable}{B1}

\begin{table*}
\caption{\label{tab:result2_SD} Standard deviation of the hyperuniformity detection discussed in section~\ref{sec:machinelearning}.
VR: Vietoris-Rips, \v{C}: \v{C}ech}
    \vspace{10pt}
\centering
\footnotesize
\begin{tabular}{@{}l||lllllll}
         \hline
         \diagbox{\bf Features}{\bf ML Model} & \makecell{linear SVM} & \makecell{SVM\\ with RBF} & \makecell{Pruned Tree} & \makecell{Random \\Forest} & \makecell{Boosted Tree} & \makecell{Simple\\ Neural \\Network} & \makecell{Deep\\ Neural \\Network} \\
         \hline    
         \hline    
         \vspace{-4pt}
         \\
         Adcock Algebraic & \makecell{VR: 0.78\%\\ \v{C}: 0.78\% }& \makecell{VR: 1.26\%\\ \v{C}: 0.99\% }& \makecell{VR: 2.11\%\\ \v{C}: 0.78\% }& \makecell{VR: 1.56\%\\ \v{C}: 0.78\% }& \makecell{VR: 1.78\%\\ \v{C}: 0.78\% }& \makecell{VR: 1.27\%\\ \v{C}: 2.54\% }& \makecell{VR: 1.82\%\\ \v{C}: 0.62\% }\\[11pt]
         Kali\v{s}nik Algebraic & \makecell{VR: 0.62\%\\ \v{C}: 0.78\% } & \makecell{VR: 0.86\%\\ \v{C}: 0.51\% } & \makecell{VR: 1.07\%\\ \v{C}: 2.68\% } & \makecell{VR: 1.77\%\\ \v{C}: 1.41\% } & \makecell{VR: 1.87\%\\ \v{C}: 1.90\% } & \makecell{VR: 0.81\%\\ \v{C}: 0.72\% } & \makecell{VR: 1.54\%\\ \v{C}: 2.02\% }\\[11pt]
         Binning $n=9$ & \makecell{VR: 1.93\%\\ \v{C}: 1.84\% } & \makecell{VR: 0.84\%\\ \v{C}: 1.15\% } & \makecell{VR: 1.10\%\\ \v{C}: 1.26\% } & \makecell{VR: 0.57\%\\ \v{C}: 1.57\% } & \makecell{VR: 0.80\%\\ \v{C}: 1.54\% } & \makecell{VR: 1.14\%\\ \v{C}: 1.99\% } & \makecell{VR: 0.67\%\\ \v{C}: 2.16\% } \\[11pt]
         Binning $n=15$ &\makecell{VR: 2.11\%\\ \v{C}: 2.33\% } &\makecell{VR: 1.25\%\\ \v{C}: 0.99\% } &\makecell{VR: 1.24\%\\ \v{C}: 2.27\% } &\makecell{VR: 1.19\%\\ \v{C}: 1.16\% } &\makecell{VR: 0.81\%\\ \v{C}: 0.74\% } &\makecell{VR: 0.47\%\\ \v{C}: 1.15\% } &\makecell{VR: 1.67\%\\ \v{C}: 1.39\% } \\[11pt]
         Binning $n=23$ &\makecell{VR: 1.60\%\\ \v{C}: 1.15\% } &\makecell{VR: 0.94\%\\ \v{C}: 0.96\% } &\makecell{VR: 1.30\%\\ \v{C}: 2.46\% } &\makecell{VR: 1.22\%\\ \v{C}: 0.91\% } &\makecell{VR: 0.61\%\\ \v{C}: 1.35\% } &\makecell{VR: 2.32\%\\ \v{C}: 1.67\% } &\makecell{VR: 1.73\%\\ \v{C}: 1.35\% }\\[11pt]
         Wasserstein distance& \makecell{VR: 0.72\%\\ \v{C}: 0.41\% } & \makecell{VR: 1.21\%\\ \v{C}: 1.55\% } & \makecell{VR: 2.93\%\\ \v{C}: 1.57\% } & \makecell{VR: 2.00\%\\ \v{C}: 0.53\% } & \makecell{VR: 1.04\%\\ \v{C}: 0.81\% } & \makecell{VR: 1.29\%\\ \v{C}: 1.97\% } & \makecell{VR: 1.91\%\\ \v{C}: 1.54\% }\\[11pt]
         \hline
    \end{tabular}
\end{table*}
\normalsize

\section{Definitions of the simplicial complexes}
\label{app:simplicial-complexes}
A simplicial complex is a collection $C$ of simplices, i.e., of points, segments, triangles, tetrahedron, and higher dimensional counterparts, with the condition that the faces of a simplex in $C$ also belong to $C$. Mathematically, a $n$-dimensional simplex with vertices $p_0, p_1, \dots, p_n$ is the convex hull of $\{p_i\}_{i=0}^n$, i.e. the smallest convex set containing $p_0, p_1, \dots, p_n$. Accordingly, a simplex is simply represented by its vertices $\{p_i\}_{i=0}^n$, the number of vertices directly implying the dimension of the simplex. 
A filtration of simplicial complexes is a collection $C_t$ for $t\in \R^{\geq 0}$ with the condition $s<t \Rightarrow C_s\subseteq C_t$.\par
For a given set of points $P$, we can define the \textit{Vietoris-Rips} ${\rm VR}_t$ and the \textit{\v{C}ech} $\rm \check C_t$ filtrations of simplicial complexes. Both are defined based on balls $B_t(p)$, i.e. ball of radius $t$ centered at $p$, with $p$ a point of $P$. Then, the two filtrations are defined as follows:
\begin{align*}
    \{p_i\}_{i=0}^n \in &{\rm VR}_t(P) \\&\Leftrightarrow \forall p_j, p_k \in \{p_i\}_{i=0}^n,\; B_t(p_j)\cap B_t(p_k)\neq \emptyset,
\end{align*}
or in words, a simplex belongs to ${\rm VR}_t(P)$ if all its edges have a distance smaller or equal to $2t$ and
$$\{p_i\}_{i=0}^n \in {\rm \check C}_t(P) \Leftrightarrow \bigcap_{i=0}^n B_t(p_i)\neq \emptyset$$
or equivalently, a simplex belongs to $\check C_t(P)$ if it can be contained within a ball of radius $t$. 
The main difference between the two considered approaches (in 2D) is that the Vietoris-Rips persistent homology will generate fewer 1-dimensional holes than the \v{C}ech. Indeed, for three given points $p_0, p_1, p_2$ and for all $t$, ${\rm VR}_t$ cannot contain the three edges $(p_0, p_1)$, $(p_0, p_2)$ and $(p_1, p_2)$ without containing the triangle $(p_0, p_1, p_2)$, while this is possible for certain $t$-values with the \v{C}ech filtration, given the triangle $(p_0, p_1, p_2)$ is non-sliver.

\section{Standard deviation of the accuracy rates by ML models}\label{app:standard deviation}
For completeness, the standard deviation of the results presented in Section \ref{sec:machinelearning}, Table \ref{tab:result2_SD}, is provided. These values are derived from conducting five separate training and testing processes, each using different datasets for every combination of feature and machine learning models. We recall that the accuracy values reported in Table \ref{tab:result2} represent the mean accuracy obtained across these five repetitions.

\section{Estimation of $H$ with Neural Network regression}

\label{app:estimationH}
While we focused in section~\ref{sec:machinelearning}
on classification between hyperuniform and non-hyperuniform patterns given a threshold on the parameter $H$, this section shows that the actual value of parameter $H$ can be estimated from the persistence diagram addressing a regression problem.

\begin{figure}[h]
     \centering
     \includegraphics[width=\linewidth]{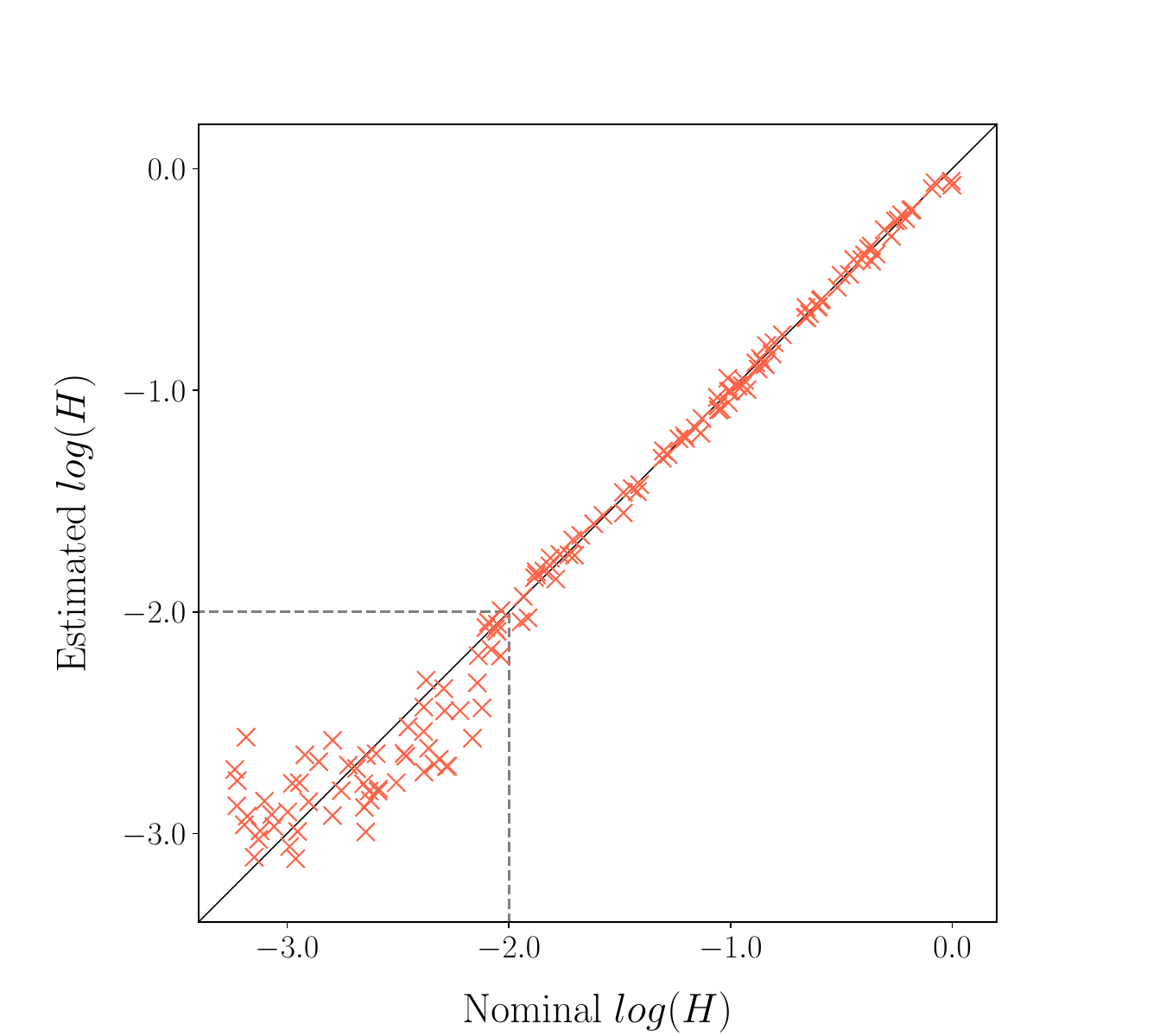}
     \vspace{-15pt}
     \caption{\label{fig:regression_H} 
     Estimation of the parameter $H$ via a neural network regression for fixed $\alpha=10$ and $K=45\pi$. Each cross represents one estimation. The accuracy is very good for nominal values higher than $10^{-2}$ while larger deviations are observed below this threshold, which is consistent with observations made in \cite{salvalaglio2024persistent}. This threshold is indicated with dashed grey lines. 
 }
 \end{figure}

It is well known in ML studies that regression generally necessitates larger training data than classification. For the following, we consequently simplify our problem by considering solely patterns with fixed $\alpha$ and $K$, but with a random $H$ value between $5\cdot10^{-4}$ and $1$ (i.e., not restricted to the values in \eqref{eq:valuesofparam}). Using the feature derived from the Wasserstein distance presented in \ref{sec:mlmodel_features_wd}, a neural network with three hidden layers of respectively $128$, $128$ and $256$ neurons is trained over $584$ realizations of such patterns with $\alpha=10$ and $K=45\pi$. This neural network was implemented with the function \textit{MPLRegressor} of the python library \textit{sklearn}. After testing the model over $146$ realizations, the accuracy of the regression is effectively good, as summarized in figure~\ref{fig:regression_H}. It is noteworthy that larger deviations are observed for a nominal $H$ smaller than $10^{-2}$: this remark is consistent with the fact that the local geometry of arrangements is only barely distinguishable for $H<10^{-2}$, as mentioned in \cite{salvalaglio2024persistent} and recalled in section \ref{sec:hu}.

\section{Details of the generation of point pattern from persistence diagrams}
\label{app:pattern-from-homology}

The method outlined in section~\ref{sec:reconstruction} to generate a pattern approximating a given $h_1$ diagram is further detailed in this appendix. It uses four basic constructions starting from a given edge $AB$ and a selected $h_1$ point also illustrated in figure~\ref{fig:constructions}. They are defined as follows.

\textit{First construction (C1):} Let $AB$ be a given edge and $l_{AB}$ its length. Let $p=(b,d)$ be an arbitrary selected $h_1$ point, with $2b>l_{AB}$. Let $A$ be the \textit{marked point} (see the role of the marked point in the algorithm \ref{alg:general}). Construct a circle of radius $d$ containing $A$ and $B$, then try to construct the point $C$ on the circle such that $l_{BC}=2b$ and $ABC$ is non-sliver; see figure~\ref{fig:constructions}(a).

\textit{Second construction (C2):} Let $AB$ be a given edge and $l_{AB}$ its length and let $p=(b,d)$ be an arbitrary selected $h_1$ point, with $2b>l_{AB}$. Let $A$ be the marked point. Construct a circle of radius $d$ such that both $A$ and $B$ belong to it. Then, construct a point $C_0$ on the circle such that $l_{BC_0}=2b$ and $l_{BC_0}\leq l_{AC_0}$. Then, divide the arc $AC_0$ to which $B$ does not belong into a number $k$ of points $C_i$, such that:
$$l_{C_0C_1}= \dots= l_{C_i C_{i+1}}= \dots= l_{C_kA}<l_{BC_0}$$
The number $k$ is chosen to be the smallest for which this condition is achieved; see figure~\ref{fig:constructions}(b).

\textit{Third construction (C3):} In opposition to the two first constructions, the $h_1$ point is not selected arbitrarily. Instead, this construction searches for a $h_1$ point satisfying given conditions. Let $AB$ be a given edge belonging to a triangle $ABC$ and $l_{AB}$ its length, and let $r$ be the radius of the circumcircle of $ABC$. Let A be the marked point. Admit furthermore that $AB$ is the longest edge of $ABC$. Without restriction of generality, among all the remaining $h_1$ points, search one with coordinates $(\bar b,\bar d)$ such that $2\bar b<l_{AB}$, $l_{AC}<2\bar b$, $l_{BC}<2\bar b$ and $r<\bar d$. Then, if possible, construct on $AB$ with this $h_1$ point using C1, else using C2; see figure~\ref{fig:constructions}(c).

\textit{Fourth construction (C4):} Let $AB$ be a given edge and $l_{AB}$ its length and let $p=(b,d)$ be an arbitrary selected $h_1$ point. Let $A$ be the marked point. Construct a right triangle $ABC$, whose hypotenuse is $AB$ and with $l_{AC}=2b$. Then, use C1 on $AC$ with the $h_1$ point $p$; see figure~\ref{fig:constructions}(d).

These constructions are used recursively in the general algorithm \ref{alg:general} to approach the targeted $h_1$ persistence diagram. It exploits two subalgorithms: the triangle-construction algorithm \ref{alg:construction} where one of the procedures C1--C4 above is applied, and the closure algorithm \ref{alg:closing} to obtain a point surrounded by triangles illustrated in figure~\ref{fig:constructions}(f). The following auxiliary functions are exploited in the algorithms: i) let $a$ be a non-constructed point, and let $\{T_i\}$ the list of triangles it belongs to, then $\theta(a)$ is the sum of all the angles (in radians) of $\{T_i\}$ for which $a$ is the angle vertex; ii) $\eta(T)$ which returns the $h_1$ point corresponding to an isolated non-sliver triangle $T$, i.e. $\eta(T)=(L/2, r)$ with $L$ the longest edge length of $T$ and $r$ its circumradius.

This proposed method is computationally efficient. On a computer with $8$ CPUs Intel(R) Xeon(R) E31275 at 3.40GHz under Debian exploitation and for the pattern size we illustrated in figure~\ref{fig:produced_patterns}, the computation time is around 10 seconds. It increases linearly with the number of $h_1$ points, and consequently linearly with the number of points in the pattern.

\begin{algorithm}
\caption{General method}\label{alg:general}
\footnotesize
\textbf{Input:} persistence diagram ($h_1$ points)\\
    -- Set the lists $L_{\rm H}$ of all targeted $h_1$ points, $L_{\rm P}$ of ``non-constructed" points containing (0,0) only, $L_{\rm C}$ (empty) of ``constructed" points.\\
    \While {First triangle is not created}{
        Select randomly an element $p_1=(b,d)$ of $L_{\rm H}$ \\
        \If{an isosceles triangle with two sides of length $2b$, and one less than $2b$ can have a circumradius of $d$
        }
        {
            -- Construct this triangle (a vertex is the origin)\\
            -- Add the new points to $L_{\rm P}$\\
            -- Remove $p_1$ from $L_{\rm H}$
            }
        }
    \While{$L_{\rm H}$ is not empty}{
        -- Select the \textit{marked point} $P$$\in$$L_{\rm P}$ as the nearest point to (0,0) \\
        -- Select one edge incident to $P$, belonging to only one triangle, preferably the largest in its triangle. Let $l_e$ be its length and $T$ its neighboring triangle\\
        -- Select randomly an element $p=(b,d)$ of $L_H$ \\
        -- Apply \textbf{algorithm \ref{alg:construction}}. \textit{Input}: $P$, $l_e$, $T$, $p$. \textit{Output}: point(s) $x_i$, $p_{\rm res} \in L_{\rm H}$\\
        \uIf{ $\theta(P)<\frac{3\pi}{2}$}{
            Remove $p_{\rm res}$ from $L_{\rm H}$ and add $x_i$ to $L_{\rm P}$
        }
        \Else{
            -- Revert the application of \textbf{algorithm \ref{alg:construction}} and apply \textbf{algorithm \ref{alg:closing}}. \textit{Input}: $P$. \textit{Outputs}: points $x_i$, $h_1$ points $p_j$\\
            -- Remove $p_j$ from $L_{\rm H}$ and add $x_i$ to $L_{\rm P}$\\
            -- Remove $P$ from $L_{\rm P}$ and add it to $L_{\rm C}$
        }
    }
\textbf{Output:} points listed in $L_{\rm C}$ and $L_{\rm H}$
\end{algorithm}
\normalsize

\begin{algorithm}
\caption{The triangle-construction algorithm}\label{alg:construction}
\footnotesize
\textbf{Input}: $P$, $l_e$, $T$, $p$, used for construction methods C1--C4\\
    \uIf{$l_e>2b$}{
        \uIf{C3 successful}{
            -- \textbf{Output}: Point $x$ generated by \textit{C3}, $p_{\rm new} \in L_{\rm H}$ as selected within $C3$. 
        }
        \Else{
            -- \textbf{Output}: Points $x_i$ generated by \textit{C4}, $p$
        }
    }
    \uElseIf{$2b<d$}{
    -- \textbf{Output}: Points $x_i$ generated by \textit{C2}, $p$
    }
    \Else{
        \uIf{C1 fails (triangle is sliver)}{
        -- \textbf{Output}: Points $x_i$ generated by \textit{C2}, $p$
        }
        \Else{
            \uIf{C1 fails (length of largest edge of triangle is~$>2b$)}{
                -- \textbf{Output}: Points $x_i$ generated by \textit{C2}, $p$
            }
            \Else{
            -- \textbf{Output}: Point $x$ generated by \textit{C1}, $p$
            }
        }
    }
    
\end{algorithm}
\normalsize

\begin{algorithm}[htbp!]
\caption{The closing procedure algorithm}\label{alg:closing}
\footnotesize
    \textbf{Input}: $P$\\
    -- Let $AP$ and $BP$ the two edges incident to $A$ and belonging to only one triangle. Let $T$ be the triangle of $AP$\\
    \uIf{$ABP$ is not sliver}{
        -- Construct triangle $ABP$\\
        -- \textbf{Output}: No point, $p \in L_{\rm H}$ closest to $\eta(ABP)$
    }
    \Else{
        \For{$\{p_i\}_1^n \in L_{\rm H}$ ($n$, tunable)}{
            -- Apply \textbf{algorithm} \ref{alg:construction}. \textit{Input}: $P$, $AP$, $T$, $p_i$. \textit{Output}: point(s) $\{x_j\}$, $p_{\rm res} \in L_{\rm H}$\\
            -- Denote $A'_i$ the only point in $\{x_j\}$ s. t. $PA'_i$ belongs to only one triangle\\
            \If{$A'_iBP$ is sliver or intersects other triangles}{
                -- Continue to $p_{i+1}$
            }
            -- Among all elements in $L_{\rm H}\setminus\{p_i\}$, let $p'_i$ be the closest to $\eta(A'_iBP)$ and $\mu_i$ their distance\\
            Revert last application of \textbf{algorithm} \ref{alg:construction}
        }
        \uIf{There is at least one non-sliver and non-intersecting $A'_iBP$}{
            -- Let $k$ be the index for which $\mu_k$ is minimal among $\mu_i$
            -- Apply \textbf{algorithm} \ref{alg:construction}. \textit{Input}: $P$, $AP$, $T$, $p_k$. \textit{Output}: point(s) $\{x_j\}$, $p_{\rm res} \in L_{\rm H}$\\
            -- Construct $A'_kBP$ with $A'_k$ the only point in $\{x_j\}$ s. t. $PA'_i$ belongs to only one triangle\\
            -- \textbf{Output}: point(s) $\{x_j\}$, $p_{\rm res}$ and $p'_k$
        }
        \Else{
            Let $C$ be the point on segment $AB$ such that $(AB)$ and $(PC)$ are perpendicular\\
            \uIf{$C$ is too close from $A$, $B$ or $P$ (tunable)}{
                -- Construct triangle $ABP$\\
                -- \textbf{Output}: No point, no $L_{\rm H}$ elements
            }
            \Else{
                Construct triangles $APC$ and $BPC$\\
                -- \textbf{Output}:  $C$, no $L_{\rm H}$ elements
            }
        }
    }
\end{algorithm}


\section*{References}
\bibliographystyle{iopart-num-mod} 

\providecommand{\noopsort}[1]{}\providecommand{\singleletter}[1]{#1}%
\providecommand{\newblock}{}

\end{document}